\documentstyle[pra,preprint,aps,epsf]{revtex}

\newcommand{\etal}{{\em et~al.\/}}

\newcommand{\opencircle}{\mbox{\Large$\circ\,$}}  
\newcommand{\opensquare}{\mbox{$\rlap{$\sqcap$}\sqcup$}}
\newcommand{\opentriangle}{\mbox{$\triangle$}}
\newcommand{\opentriangledown}{\mbox{$\bigtriangledown$}}
\newcommand{\opendiamond}{\mbox{$\diamondsuit$}}
\newcommand{\fullcircle}{\mbox{{\Large$\bullet\,$}}} 
\newcommand{\fullsquare}{\,\vrule height5pt depth0pt width5pt}

\begin{document}
\draft

\title{Integral cross sections for electron scattering by \\ 
ground state Ba atoms}

\author{D.~V.~Fursa$^\dag$\thanks{electronic address: dmitry.fursa@flinders.edu.au}, 
S.~Trajmar$^\ddag$, I.~Bray$^\dag$, I.~Kanik$^\ddag$, \\
 G.~Csanak$^\S$ , R.E.H.~Clark$^\S$, and J.~{Abdallah~Jr.$^\S$}
}
\address{$^\dag$ The Flinders University of South Australia,
G.P.O. Box 2100, Adelaide 5001, Australia\\
$^\ddag$ Jet Propulsion Laboratory, California Institute of Technology, 
Pasadena, CA, USA\\
$^\S$ Los Alamos National Laboratory, University of California, Los Alamos, 
NM 887545, USA
}

\date{\today}
\maketitle
\begin{abstract}
We have used the convergent close-coupling method and a unitarized 
first-order many-body 
theory to calculate integral cross 
sections for elastic scattering and momentum transfer, for
excitation of the 5d$^2$~$^1$S, 6s6p~$^1$P$_1$, 6s7p~$^1$P$_1$, 6s8p~$^1$P$_1$, 
6s5d~$^1$D$_2$, 5d$^2$~$^1$D$_2$, 
6s6d~$^1$D$_2$, 6p5d~$^1$F$_3$, 6s4f~$^1$F$_3$, 6p5d~$^1$D$_2$, 
6s6p~$^3$P$_{0,1,2}$, 6s5d~$^3$D$_{1,2,3}$, and 6p5d~$^3$D$_2$ states,
for ionization and for total scattering
by electron impact on the ground state of barium at incident electron energies 
from 1 to 1000~eV. These results and all available experimental data 
have been combined to
produce a recommended set of integral cross sections. 

\end{abstract}

\pacs{34.80.Bm, 34.80.Dp}

\newpage

\section{Introduction}
A great deal of interest and need has developed in recent years for
electron collision  
cross sections involving Ba atoms. In the applications area, these
cross sections 
are needed for modelling the behavior of Ba vapor lasers 
\cite{MBCP95,MBP97a,MBP97b,MBP97c},
discharge lamps \cite{Bhattacharya89}, plasma switches \cite{YR92},
and various  planetary ionospheres 
\cite{Westcott80,Simons81,Winske88,Chapman89,Shuk.Szuszgzewier91,Westcott.etal93}, 
where Ba has often been used as a trace element 
for diagnostic purposes. On the academic side, benchmark laboratory
cross sections  
are needed for testing various theoretical approximations and
calculational methods hoping to predict these cross sections.

The experimental data base, available at the present time, is rather
limited both in the electron impact energy range and the scattering
channels. Line emission cross sections for the ($6s6p~^1P_1
\rightarrow 6s^2~^1S_0$) at 5535~{\AA} [$Q_{emiss}$($6s6p~^1P_1$)]
were determined by Chen and Gallagher \cite{CG76} in the 2.3 to 1497.0
eV impact energy range. They claimed an uncertainty of $\pm$5~\%.
Since the $6s6p~^1P_1$ level decays predominantly (99.7\%) to the
ground state, the measured line emission cross sections are equivalent
(within the experimental error limits) to the apparent $6s6p~^1P_1$
level excitation cross sections [$Q_{App}$($6s6p~^1P_1$)] and they
differ from the electron impact excitation cross sections
[$Q$($6s6p~^1P_1$)] by the cascade contributions. (See e.g. Trajmar
and Nickel \cite{TN92} for the definitions of these cross sections.)
Cascade corrections, only available from theory, can be applied to the
data of Chen and Gallagher and the resulting $Q$($6s6p~^1P_1$) values
represent the most reliable electron scattering cross sections
available for Ba at the present time. Jensen \etal \cite{JRT78} and
Wang \etal \cite{WTZ94} determined relative cross sections for elastic
scattering ($Q_{\rm elas}$) and momentum transfer ($Q_M$) at a few impact
energies.  Jensen \etal \cite{JRT78} also obtained some cross section
results for excitation of the $6s5d~^1D_2$ level
[$Q$($6s5d~^1D_2$)]. In these cases, the relative cross sections were
normalized by an estimated cascade correction
applied to the Chen and Gallagher $Q_{App}$($6s6p~^1P_1$) values to
obtain $Q$($6s6p~^1P_1$) values which in turn were used to normalize
$Q_{\rm elas}$, $Q_M$, and $Q$($6s5d~^1D_2$). Total ionization cross
section ($Q_i$) in the threshold to 600 eV range have been reported by
Dettmann and Karstensen \cite{DK82} and by Vainshtein \etal \cite{VORS72}
from  the threshold to 200 eV. Total electron scattering
cross sections ($Q_{\rm Tot}$) were measured by Romanyuk \etal
\cite{RSZ80} in the 0.1 to 10 eV range.

There is a larger data base available from calculations.  Elastic
scattering cross sections were calculated by Gregory and Fink
\cite{GF74} in the 100 to 1500 eV range.  (numerical solutions of the
Dirac equation), by Fabrikant \cite{Fabrikant80} at impact energies
ranging from 6 eV to 35 eV (non-relativistic close-coupling
approximation), by Yuan and Zhang from 0.01 eV to 5.0 eV 
(quasirelativistic static-exchange formalism) \cite{YZ90} and from 0.04 eV
to 150 eV (Hartree-Fock method with relativistic corrections)
\cite{YZ92}, by Szmytkowski and Sienkiewicz \cite{SS94} in the 0.2 eV
to 100 eV region (relativistic polarized-orbital approximation) and by
Kelemen \etal
\cite{KRS95} from 0.1 to 200 eV (using phenomenological complex opical
potential).  Szmytkowski and Sienkiewicz \cite{SS94} and Kelemen \etal
\cite{KRS95} as well as Gribakin \etal \cite{GGIKT92} (Hartree-Fock
approximation with correlation corrections, from zero to 2.5 eV) have
reported momentum transfer cross sections.  As far as inelastic
scattering is concerned, $Q$($6s6p~^1P_1$) results were obtained by
Fabrikant \cite{Fabrikant80} from threshold to 35 eV (non-relativistic
two-state close-coupling approximation), by Clark \etal \cite{CACK89}
from 5 eV to 100 eV (unitarized distorted-wave approximation, UDWA and
first order many-body theory, FOMBT), and Srivastava \etal
\cite{SZMS92,SMS92} from 20 to 100 eV (relativistic distorted-wave
approximation, RDWA). Srivastava \etal 
also reported $Q$($6s6p~^3P_1$)
and $Q$($6s5d~^1D_2$) and $Q$($6s5d~^3D_{1,2,3}$) values.  $Q_{\rm Tot}$
results in the 10 to 200 eV range were given by Kelemen \etal
\cite{KRS95}.  Very recently the non-relativistic convergent
close-coupling (CCC) method was applied by Fursa and Bray
\cite{FB98R,FB99} to obtain $Q_{\rm elas}$, $Q_M$, $Q$($6s6p~^3P_1$),
$Q$($6s5d~^1D_2$) and $Q_{\rm App}$($6s6p~^1P_1$) results in the 1 to 897
eV range.

The present work represents a substantial extension of CCC and UFOMBT
calculations to cover all scattering channels which we consider
important for practical applications over a wide range of impact
energies. Comparison of these theoretical results with fragmentary
experimental data allows us to recommend a reliable and consistent
cross section data set which should be satisfactory for most modelling
calculations. We found very good agreement between the CCC results and
experiment and therefore in our recommendations relied heavily on the
CCC data.

\newpage
\section{Calculational methods}

\subsection{CCC method}
The application of the CCC method to calculation of electron
scattering from barium has been discussed elsewhere, see
Refs. \cite{FB99} and \cite {FB97} for details.  Briefly, barium
target states are described by a model of two valence electrons above
an inert Hartree-Fock core. We have used configuration interaction
(CI) expansion technique to obtain barium wave functions. One-electron
orbitals used in CI expansion have been obtained by diagonalizing
Ba$^+$ Hamiltonian in a Sturmian (Laguerre) basis.  In Table
\ref{energy.levels} we compare energies for the states relevant to the
present study with experimental data and give a set of the dominant
configurations for each state.  We find a very good agreement between
our results and experiment and other accurate calculations for energy
levels and oscillator strengths \cite{FB99}.  The barium target states
obtained this way provide not only an accurate representation of the
barium discrete spectrum but allow also for square-integrable
representation of the target continuum. This allows for coupling to
the ionization channels in the scattering calculations. These
calculations use barium target states in order to perform expansion of
the total wave function and formulate a set of close-coupling
equations. These equations (for the $T$ matrix) are formulated and
solved in momentum space.

The CCC method is formulated as a purely non-relativistic theory in
both target structure and electron scattering calculations.  In order
to compare results from the non-relativistic CCC calculations with
experiment, we have used a technique essentially identical with the
transformation scheme described by Saraph \cite{Saraph72}.  
Namely, we first transform the non-relativistic CCC scattering amplitudes
$f^S_{\pi_f s_f l_f m_f,\pi_i s_i l_i m_i}$ to
the amplitudes describing transitions between fine-structure levels
$J_f$ and $J_i$,
\begin{eqnarray}
f^{\sigma_f,\sigma_i}_{\pi_f J_f M_f,\pi_i J_i M_i}(s_f l_f \gamma_f, s_i l_i \gamma_i) = \sum_{m_f,q_f,m_f,q_f,S} &&
  C^{J_f M_f}_{l_f m_f,s_f q_f} C^{S M_S}_{{1 \over2}\sigma_f,s_f q_f}  
  C^{J_i M_i}_{l_i m_i,s_i q_i} C^{S M_S}_{{1 \over2}\sigma_i,s_i q_i} \nonumber \\
  && \; f^S_{\pi_f s_f l_f m_f,\pi_i s_i l_i m_i}(\gamma_f,\gamma_i).
\label{non.rel}
\end{eqnarray}
Here $S$ is total spin, and $\pi_f$ ($\pi_i$), $s_f$ ($s_i$), $l_f$
($l_i$) and $m_f$ ($m_i$) are the final (initial) target state parity,
spin, orbital angular momentum is and its projection on the Z-axis 
of the collision frame, respectively.  The final (initial) projectile
spin projection on the Z-axis of the collision frame is indicated as
$\sigma_f$ ($\sigma_i$) , and the index $\gamma$ distinguishes states
with the same orbital angular momentum, spin and parity. The above
amplitudes are used to form amplitudes in the intermediate coupling
scheme
\begin{equation}
F^{\sigma_f,\sigma_i}_{\pi_f J_f M_f,\pi_i J_i M_i}(\beta_f,\beta_i) = 
\sum_{s_f,l_f, s_i, l_i}  \, \sum_{\gamma_f,\gamma_i} 
C^{\beta_f}_{\gamma_f} \, C^{\beta_i}_{\gamma_i} \,
f^{\sigma_f,\sigma_i}_{\pi_f J_f M_f,\pi_i J_i M_i}(s_f l_f \gamma_f,s_i l_i \gamma_i),
\label{semi.rel}
\end{equation}
where the index $\beta$ distinguishes target states with the same
total angular momentum $J$ and parity $\pi$.  We obtain mixing
coefficients $C^{\beta}_{\gamma}$ by diagonalizing the Breit-Pauli
Hamiltonian (only one-body spin-orbit term is used) in the basis of
the barium target states obtained from the non-relativistic barium
structure calculation.  Note that the dependence of the scattering
amplitudes in (\ref{non.rel}) and (\ref{semi.rel}) on the electron
spherical angles $\theta$ and $\varphi$ is implicit.

Amplitudes (\ref{semi.rel}) are used to calculate the semi-relativistic
integrated cross sections:
\begin{equation}
Q_{\rm fs} = {k_f \over 2(2J_i+1)k_i} \sum_{M_f,M_i,m_f,m_i} \int
d\Omega |F^{\sigma_f,\sigma_i}_{\pi_f J_f M_f,\pi_i J_i
M_i}(\beta_f,\beta_i)|^2. 
\label{ics}
\end{equation}
The subscript ``fs'' (fine-structure) indicates that the cross section is
calculated with an
(approximate) account of relativistic corrections.
	
	Scattering on a singlet initial state allows for significant
simplification in Eq. (\ref{ics}). Symmetry relations of the
scattering amplitudes (\ref{non.rel})
\begin{equation}
f^{\sigma_f,\sigma_i}_{\pi_f J_f M_f,\pi_i J_i M_i}(s_f l_f \gamma_f, s_i l_i \gamma_i) = 
-(-1)^{s_f}\, f^{-\sigma_f,-\sigma_i}_{\pi_f J_f M_f,\pi_i J_i M_i}(s_f l_f \gamma_f, s_i l_i \gamma_i), \; s_i = 0,
\label{sym}
\end{equation}
ensure that the singlet-triplet terms in Eq. (\ref{ics}) are zero after
summation over projectile spin magnetic sublevels $m_f$ and $m_i$. We have
also found that for the target states involved in the present study
only  one or two  
terms  in Eq. (\ref{semi.rel}) have large mixing
coefficients. Together, these allow us 
to express the cross section defined by (\ref{ics}) in terms of
the non-relativistic cross sections $Q$ 
which are obtained from the non-relativistic amplitudes (\ref{non.rel}) 
using 
Eq. (\ref{ics}). We give below decomposition of the  semi-relativistic ICS (\ref{ics})
via non-relativistic cross sections,
\begin{mathletters}
\begin{eqnarray}
Q_{\rm fs}(5d^{2}~{^1}S_0) &=& 0.9635 \, Q(5d^2~{^1}S_0) + 0.0339 \, Q(5d^2~{^3}P_0) \\ 
Q_{\rm fs}(6s6p~{^3}P_1) &=& 0.9934 \, Q(6s6p~{^3}P_1) + 0.0058 \, Q(6s6p~{^1}P) \\ 
Q_{\rm fs}(6s5d~{^1}D_2) &=& 0.9779 \, Q(6s5d~{^1}D_2) + 0.0220 \, Q(6s5d~{^3}D_2) \\ 
Q_{\rm fs}(6s5d~{^3}D_2) &=& 0.9779 \, Q(6s5d~{^3}D_2) + 0.0220 \, Q(6s5d~{^1}D_2) \\ 
Q_{\rm fs}(6s6d~{^1}D_2) &=& 0.9845 \, Q(6s6d~{^1}D_2) + 0.0136 \, Q(6s6d~{^3}D_2) \\ 
Q_{\rm fs}(5d^{2}~{^1}D_2)   &=& 0.8591 \, Q(5d^2~{^1}D_2) + 0.1292 \, Q(5d^2~{^3}P_2) \\ 
Q_{\rm fs}(6p5d~{^1}D_2) &=& 0.7774 \, Q(6p5d~{^1}D_2) + 0.2091 \, Q(6p5d~{^3}F_2) \\  
Q_{\rm fs}(6p5d~{^3}D_2) &=& 0.9878 \, Q(6p5d~{^3}D_2) + 0.0075 \, Q(6p5d~{^1}D_2)  \\ 
\label{5h}
Q_{\rm fs}(6p5d~{^1}F_3) &=& 0.9698 \, Q(6p5d~{^1}F_3) + 0.0291 \, Q(6p5d~{^3}D_3).
\end{eqnarray}
\end{mathletters}
These cross sections typically differ by less than 3\% from 
the corresponding cross sections obtained from Eq. (\ref{ics}). 
All other target states are well
described in the non-relativistic approximation.

\subsection{UFOMBT method}
The UFOMBT method used here has been discussed in general and in
particular its implementation for Ba by Clark \etal \cite{CACK89} and
Zetner \etal \cite{Zetner97}.

\section{Results and discussion}

\subsection{Line emission, apparent level excitation and electron impact 
excitation cross section for the $6s6p~^1P_1$ level}

\begin{figure}
\vspace{5cm}
\hspace{-2cm}
\epsfbox{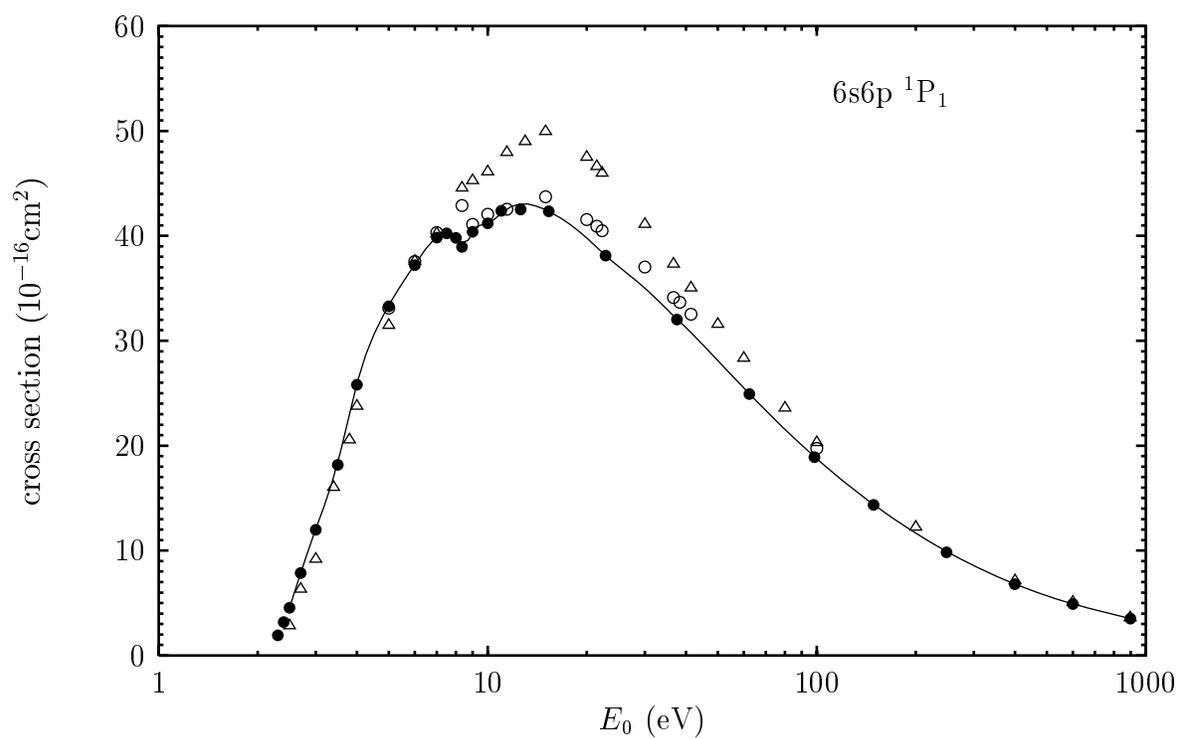}
\vspace{1cm}
\caption{Apparent 6s6p$^1$P$_1$ 
integral excitation cross sections:
\opencircle, CCC; \opentriangle, CC(55);
\fullcircle, Chen and Galagher ~\protect\cite{CG76}.
The solid line represents our recommended values.
}
\label{ics.s6P_app}
\end{figure}

\begin{figure}
\hspace{-2cm}
\epsfbox{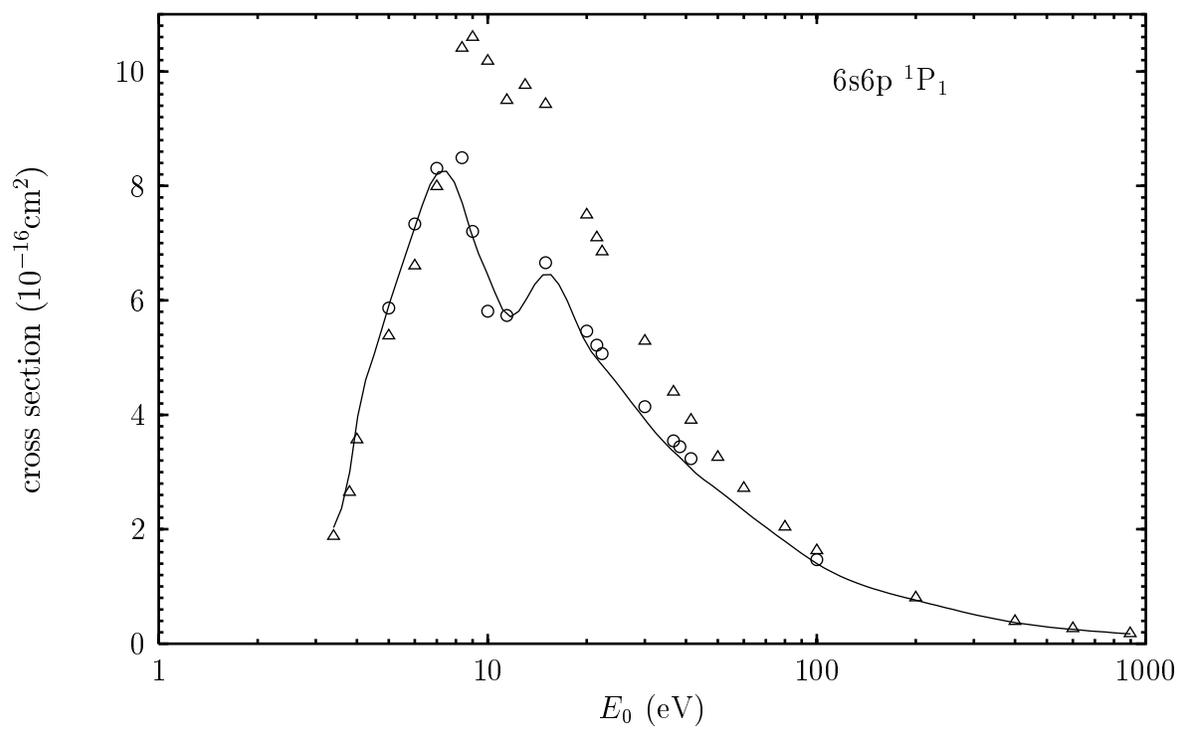}
\vspace{1cm}
\caption{\rm Cascade contribution to the 6s6p $^1$P$_1$ level apparent
excitation cross section: \opencircle, CCC; \opentriangle, CC(55).  
The solid line represents our recommended values.}
\label{ics.s6P_casc}
\end{figure}

\begin{figure}
\hspace{-2cm}
\epsfbox{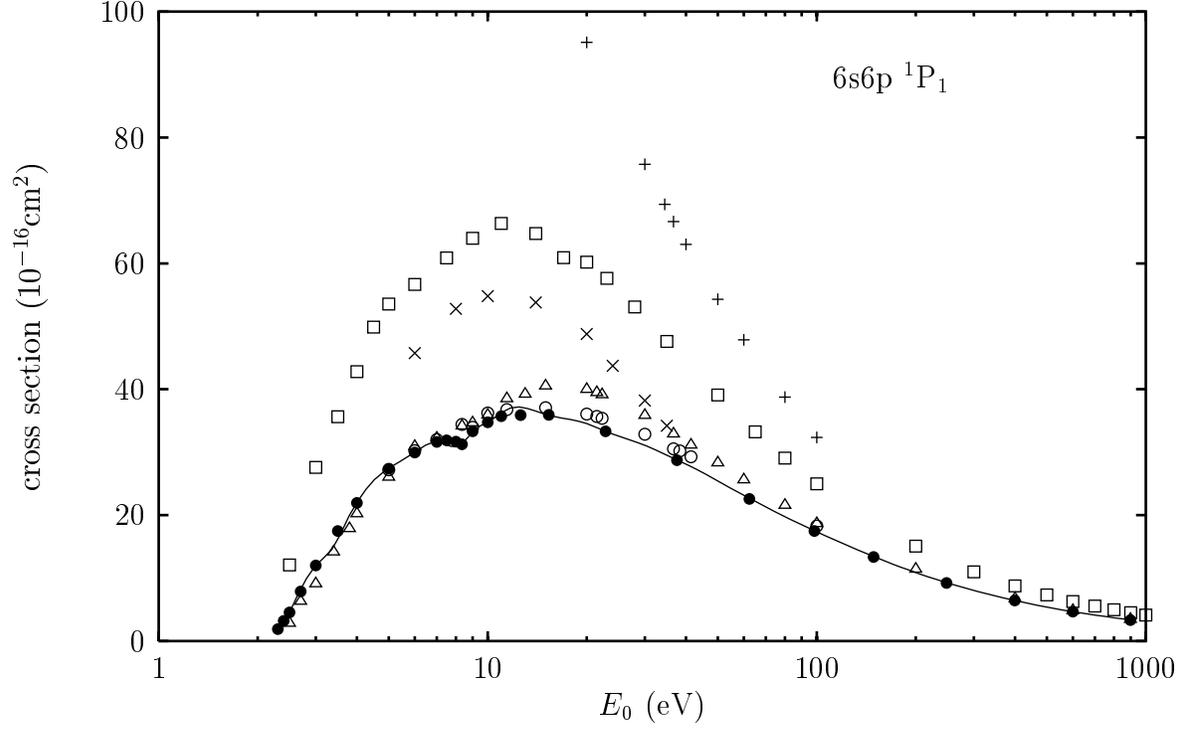}
\vspace{1cm}
\caption{Integral cross sections for excitation of the 6s6p$^1$P$_1$
level: \opencircle, CCC; \opentriangle, CC(55);\opensquare, UFOMBT; 
$\times$, CC(2) Fabrikant \protect\cite{Fabrikant80};  
+, RDWA Srivastava \etal \protect\cite{SZMS92};
\fullcircle, obtained from apparent cross section of Chen and Galagher
\protect\cite{CG76} by subtracting theoretical (CCC and CC(55))
estimate of cascade contribution. The solid line represents our
recommended values. }
\label{ics.s6P}
\end{figure}

At the present time, the most reliable electron collision cross
sections for Ba are the 5535~{\AA} line emission cross sections
[$Q_{\rm emiss}$($6s6p~^1P_1$)] associated with the radiative decay of the
electron impact and cascade populated $6s6p~^1P_1$ level to the ground
$6s^2~^1S_0$ state as measured by Chen and Galagher \cite{CG76}. The
uncertainty claimed for these cross sections is about $\pm5\%$ over
the 2.3 to 1497 eV impact energy range. As mentioned in the
Introduction, for all practical purposes these emission cross sections
are equivalent to the apparent level excitation cross sections
[$Q_{\rm App}$($6s6p~^1P_1$)] from which the electron impact excitation
cross sections [$Q$($6s6p~^1P_1$)] can be derived if proper account
for the cascade contributions can be made.  These cross sections can
be used as standards to normalize other electron collision cross
sections obtained from relative measurements. Indeed, this procedure
was followed by Jensen \etal \cite{JRT78} and Wang \etal \cite{WTZ94}
who assumed very approximate cascade contributions. A better estimate
of these cascade contributions can be made based on the CCC
calculations. We will follow here this latter procedure.  In
Fig.~\ref{ics.s6P_app} $Q_{\rm App}$($6s6p~^1P_1$ ) values measured by
Chen and Gallagher and those obtained from the CCC and CC(55)
calculations (by adding the direct and cascade contributions) are
shown.  Fig.~\ref{ics.s6P_casc} shows the calculated cascade
contribution. Chen and Gallagher have used the Bethe-Born theory to
normalize their relative measurements at high energy. They used the
value of the optical oscillator strength $f=1.59$ a.u. for the $6s^2~^1S_0$-
$6s6p~^1P_1$ transition. This value is now known more accurately,
$f=1.64$ a.u. \cite{BJLMP85}.  We, therefore, have multiplied the cross
section values given by Chen and Gallagher by the ratio of the latter
and former optical oscillator strengths.  The excellent agreement
between experiment and the CCC results gives credence to the CCC
method and some assurance that the $Q$($6s6p~^1P_1$) cross sections
from these calculations are reliable.  In Fig.~\ref{ics.s6P}, we
compare these cross sections with those obtained from the Chen and
Gallagher $Q_{\rm App}$($6s6p~^1P_1$ ) and the results obtained
from other calculational methods. As can be seen from
Fig. \ref{ics.s6P}, the calculational methods converge at higher
impact energies (above few hundreds eV) but only the CCC results can
be considered reliable at intermediate and low impact energies.  The
set of recommended cross sections are given in Table~\ref{II}. The
apparent cross sections are those of Chen and Gallagher, marginally
renormalized by multiplication by 1.03 as discussed above. The ratio
of $Q_{\rm cascade} / Q_{\rm App}$ has been evaluated using the CCC and CC(55)
results. Both recommended cascade $Q_{\rm cascade}$ and direct
$Q$(6s6p$^1$P$_1$) cross sections have been obtained from the apparent
cross sections with the utilization of the CCC $Q_{\rm cascade} / Q_{\rm App}$
ratio.

\subsection{Other inelastic scattering channels}
In all UFOMBT calculations except for the excitation of the
$6s4f~^1F_3$ and the $6p5d~^1D_2$ levels the 22 configurational basis
set described in Zetner \etal \cite{Zetner97} was used. 

\begin{figure}
\hspace{-2cm}
\epsfbox{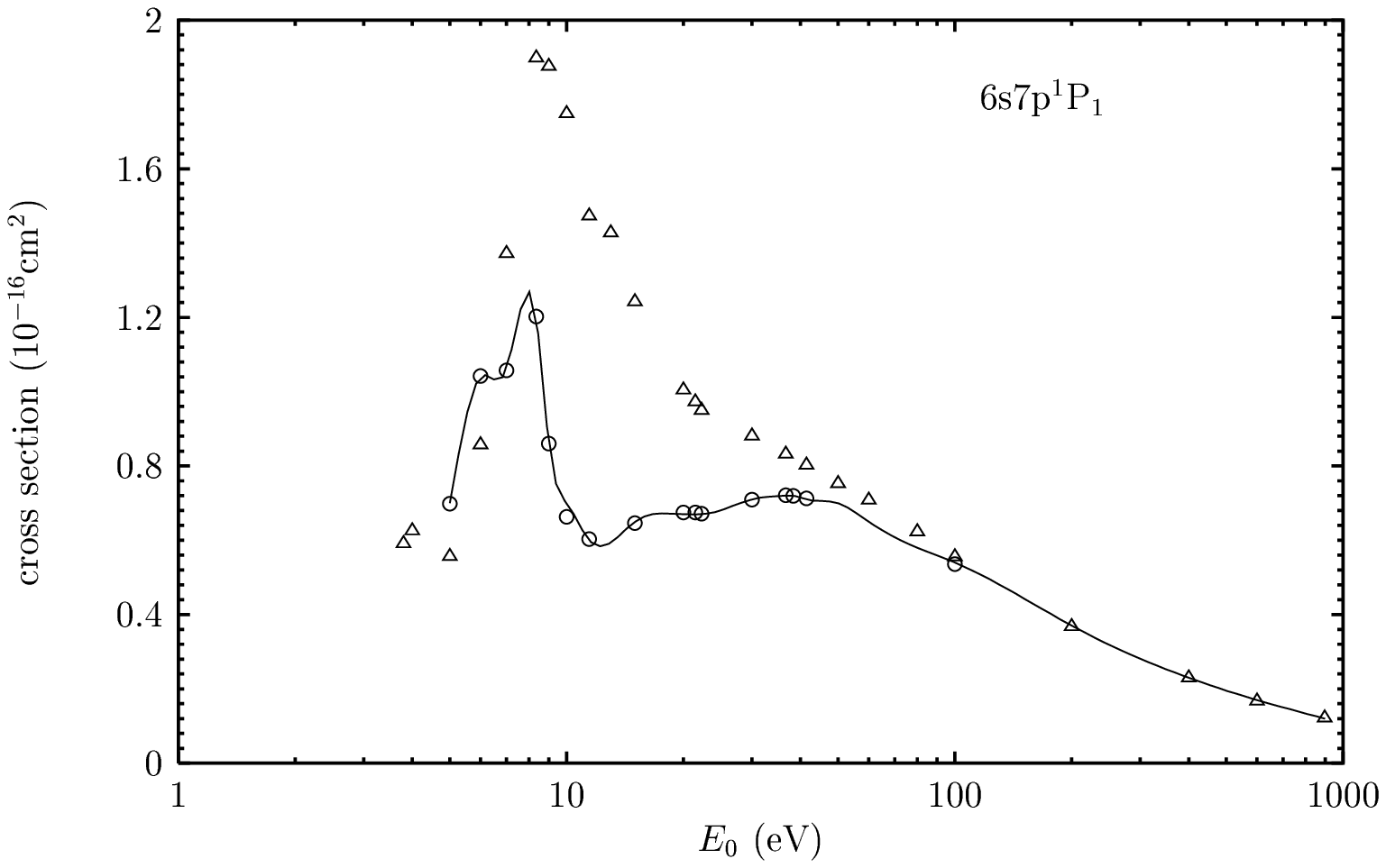}
\vspace{1cm}
\caption{Same as Fig.~\ref{ics.s6P_app} except for the 6s7p $^1$P$_1$ level.
}
\label{ics.s7P_app}
\end{figure}

\begin{figure}
\hspace{-2cm}
\epsfbox{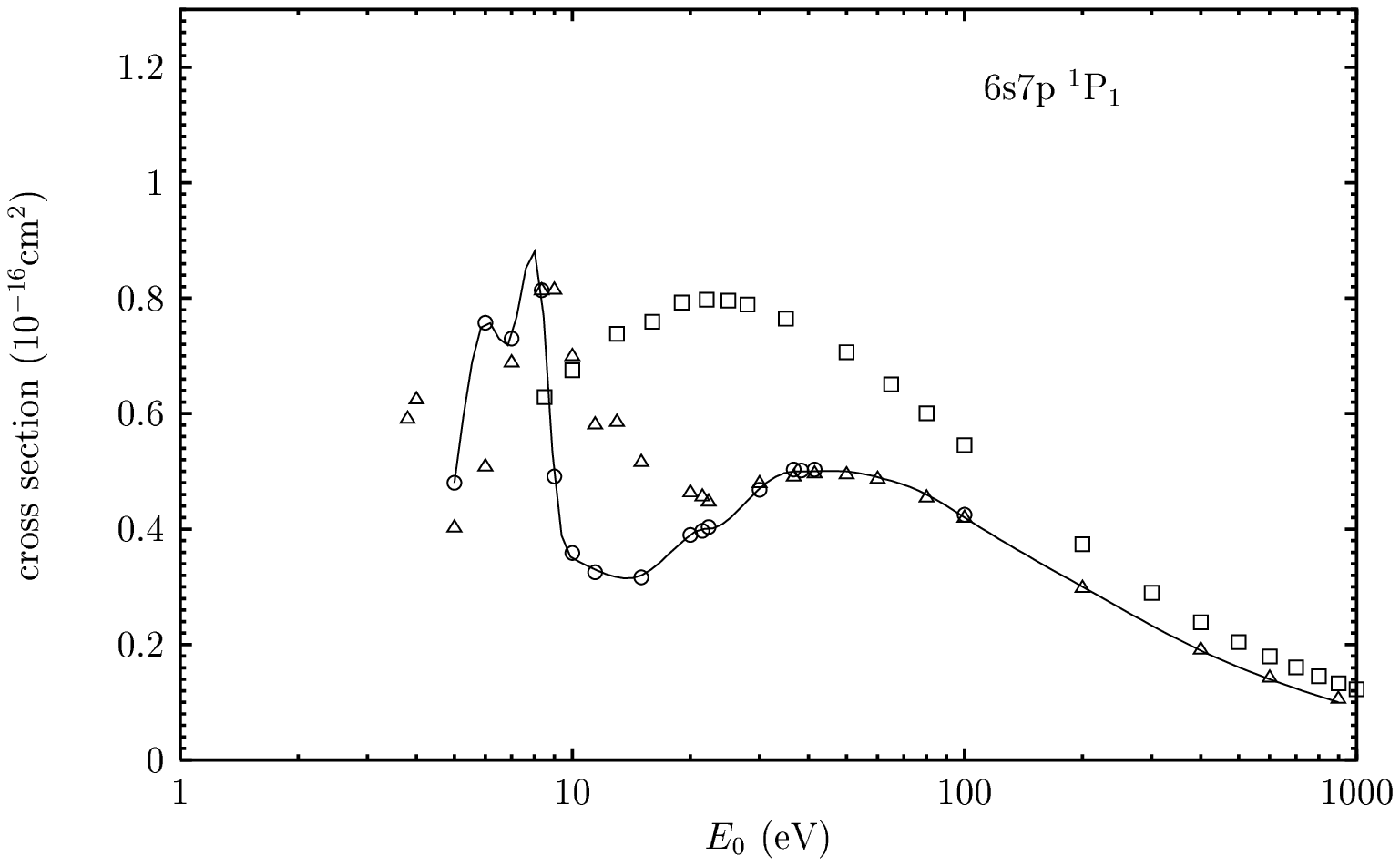}
\vspace{1cm}
\caption{Same as Fig.~\ref{ics.s6P} except for the 6s7p $^1$P$_1$ level.
}
\label{ics.s7P}
\end{figure}

\begin{figure}
\hspace{-2cm}
\epsfbox{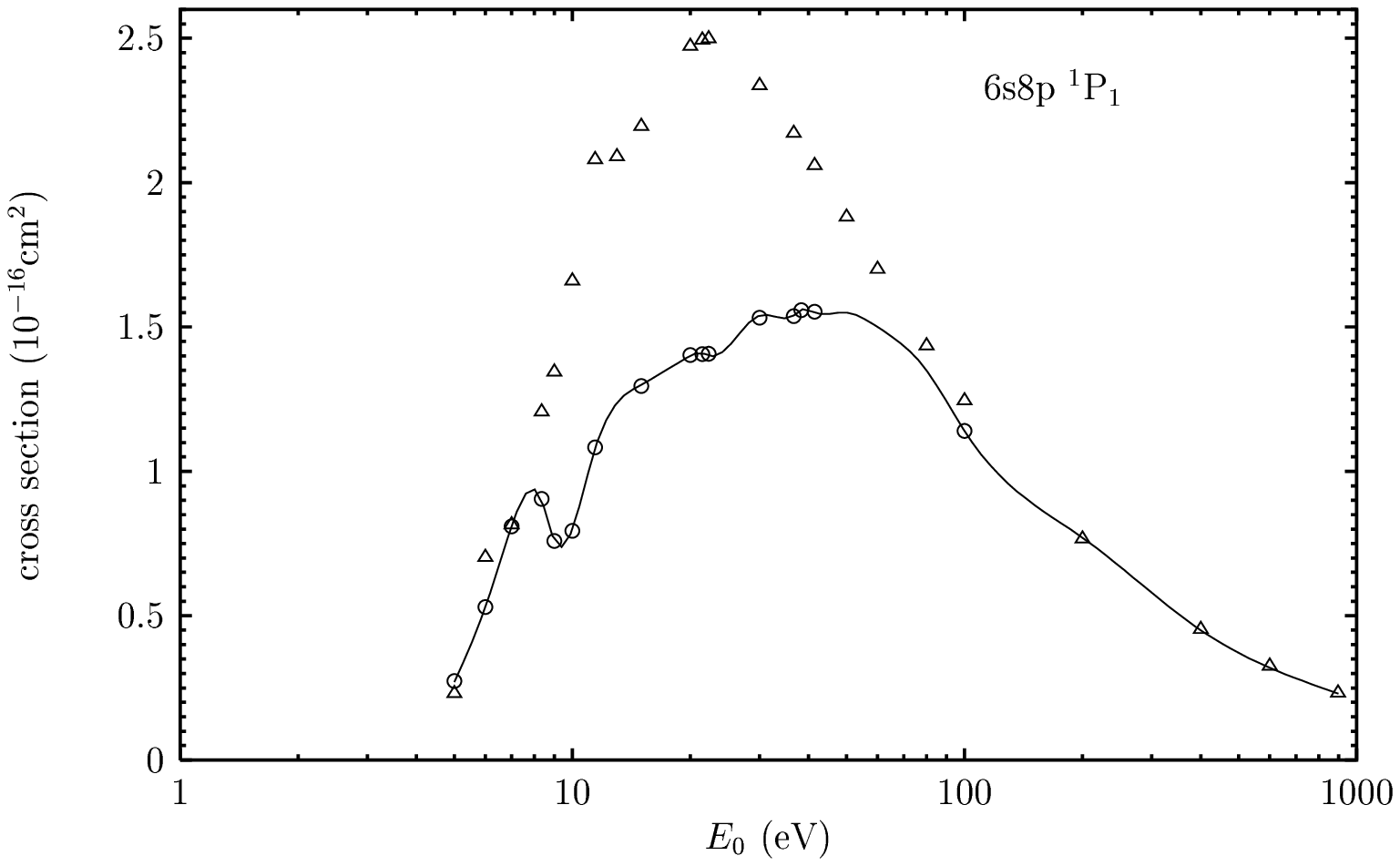}
\vspace{1cm}
\caption{Same as Fig.~\ref{ics.s6P_app} except for the 6s8p $^1$P$_1$ level.
}
\label{ics.s8P_app}
\end{figure}

\begin{figure}
\hspace{-2cm}
\epsfbox{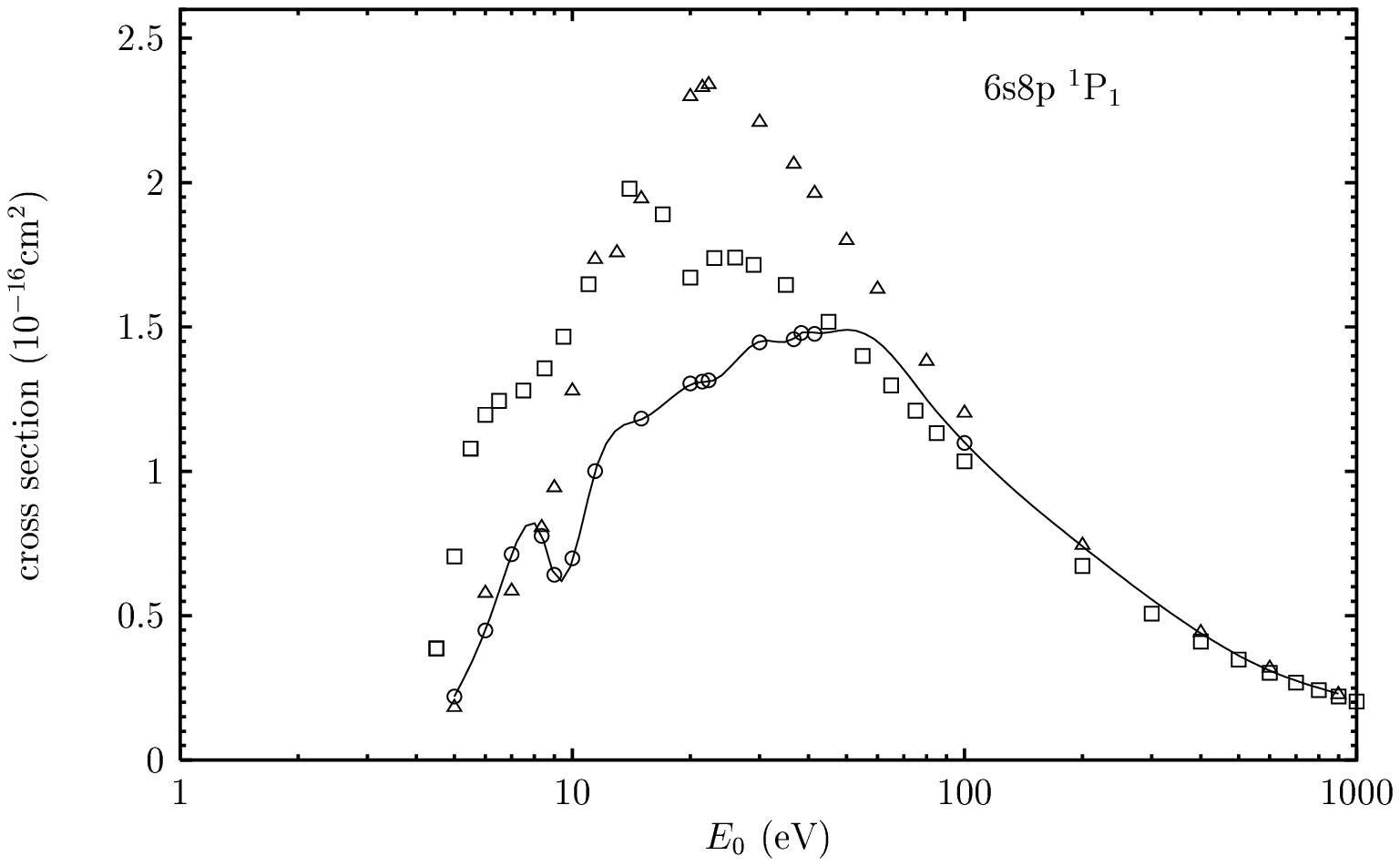}
\vspace{1cm}
\caption{Same as Fig.~\ref{ics.s6P} except for the 6s8p $^1$P$_1$ level.
}
\label{ics.s8P}
\end{figure}

Apparent level excitation and electron impact excitation cross
sections for the $6s7p~^1P_1$ and $6s8p~^1P_1$ levels, obtained from
CCC, CC(55) and UFOMBT calculations, are shown in
Figs.~\ref{ics.s7P_app}, \ref{ics.s7P} and ~\ref{ics.s8P_app},
\ref{ics.s8P}, respectively.  No experimental data or other
theoretical results are available for these excitation processes. The
recommended cross sections are listed in Table~\ref{III}.
These values correspond to the CCC results.  No recommended cross
sections are given below 5.0 eV since the present implementation of the 
CCC method is too computationally expensive to study resonance regions.

\begin{figure}
\hspace{-2cm}
\epsfbox{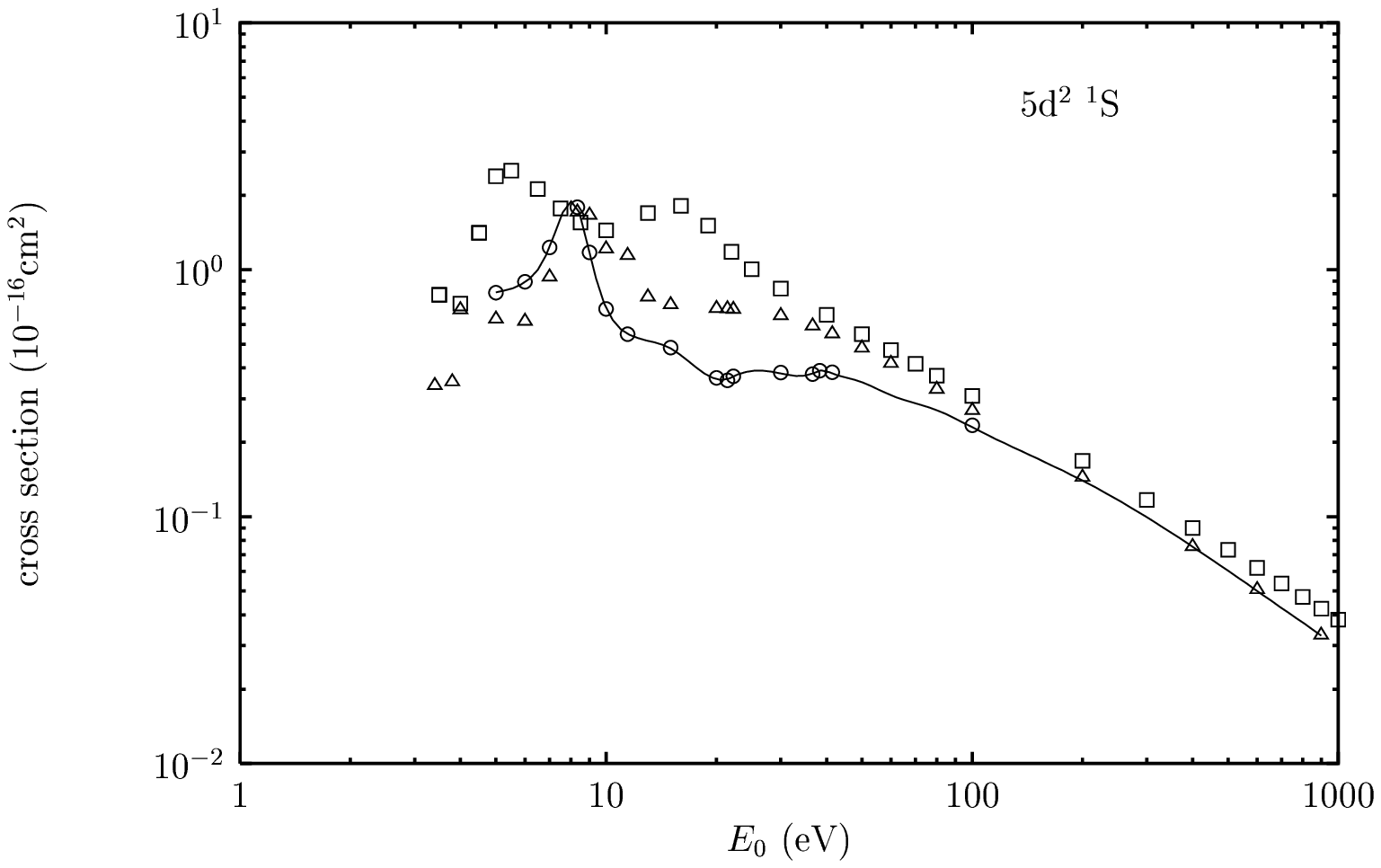}
\vspace{1cm}
\caption{Same as Fig.~\ref{ics.s6P} except for the 5d$^2~^1$S level. 
}
\label{ics.s7S}
\end{figure}

\begin{figure}
\hspace{-2cm}
\epsfbox{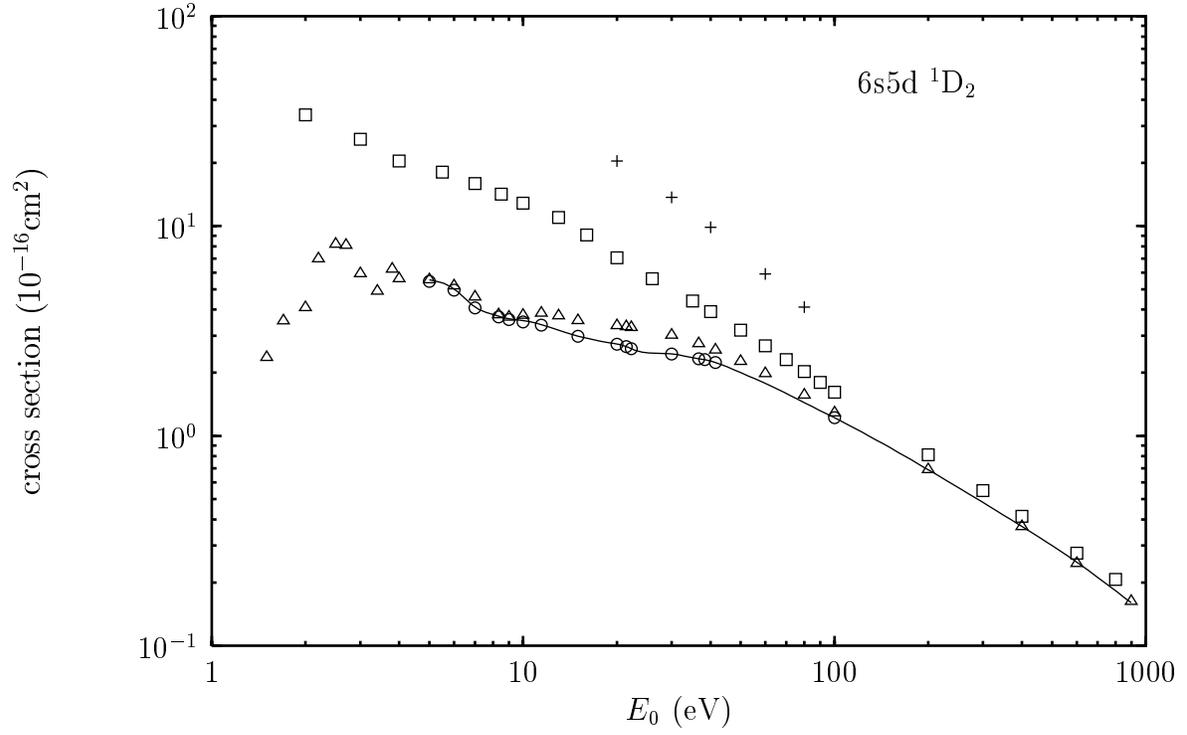}
\vspace{1cm}
\caption{Integrated cross sections for excitation of the 6s5d$^1$D$_2$ level:
\opencircle, CCC; \opentriangle, CC(55);
\opensquare, UFOMBT;
+, RDWA  Srivastava \etal \protect\cite{SMS92}.
The solid line represents our recommended values.
}
\label{ics.s5D}
\end{figure}

\begin{figure}
\hspace{-2cm}
\epsfbox{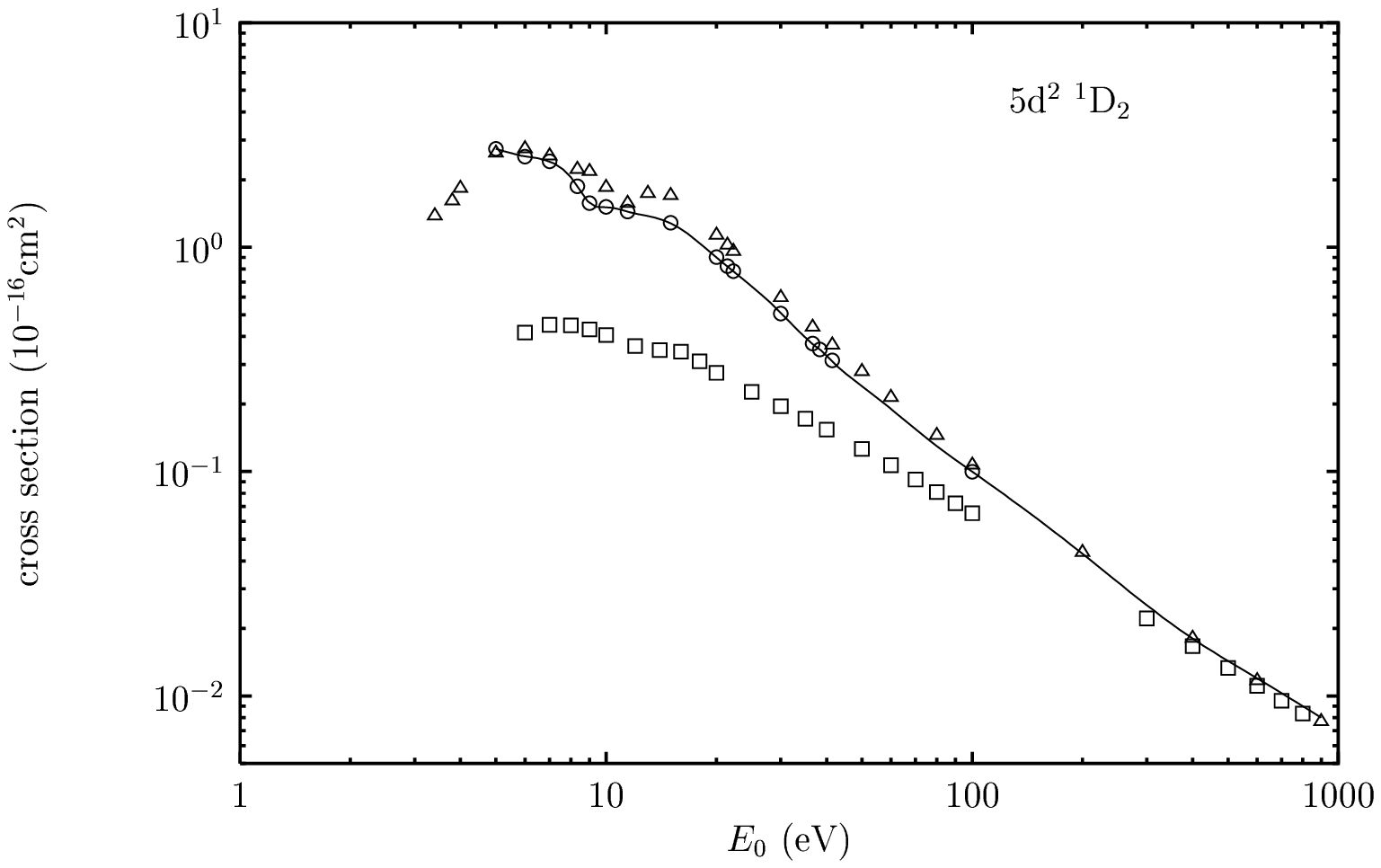}
\vspace{1cm}
\caption{Same as Fig.~\ref{ics.s5D} except for the 5d$^2$ $^1$D$_2$ level.
}
\label{ics.s6D}
\end{figure}

\begin{figure}
\hspace{-2cm}
\epsfbox{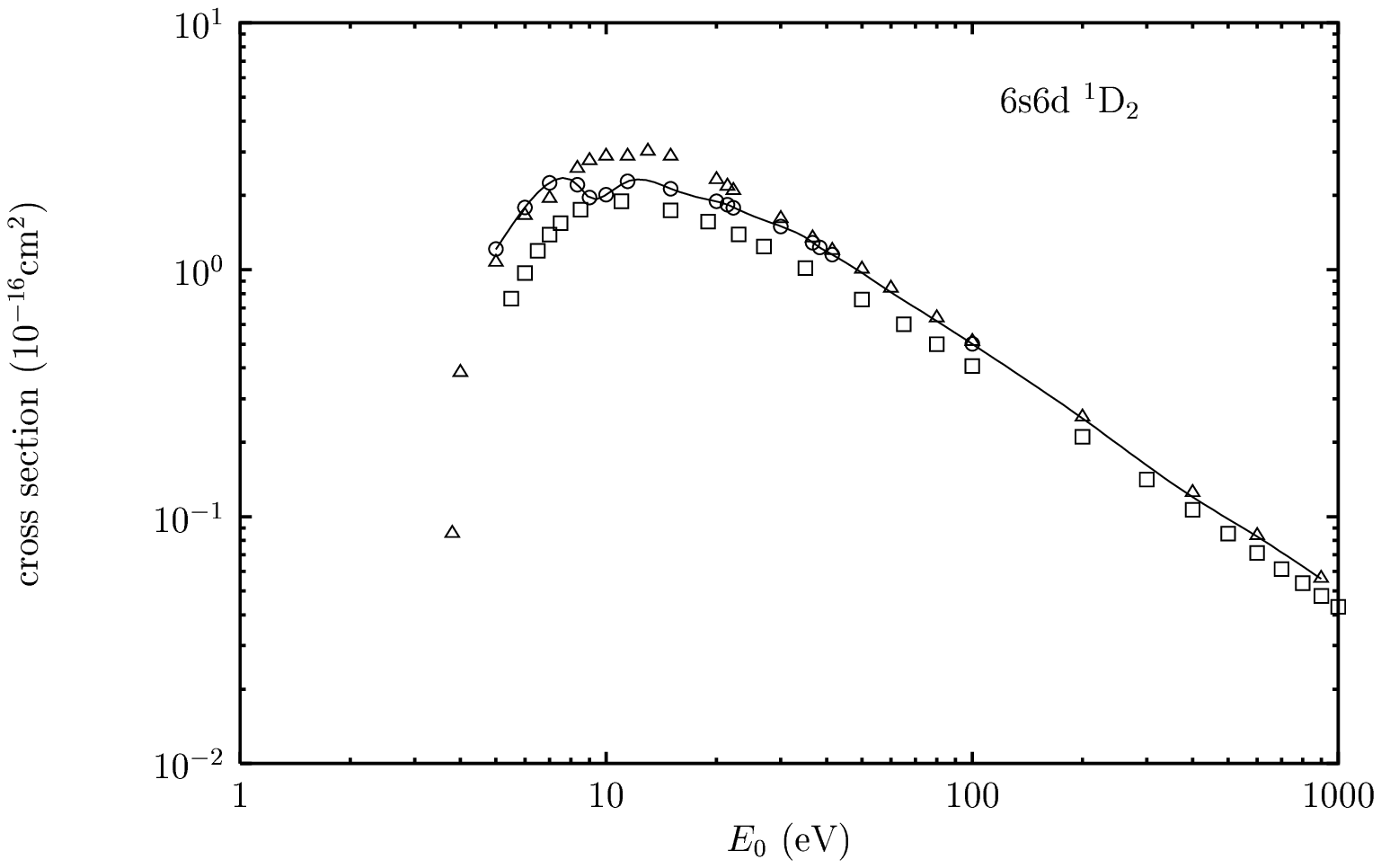}
\vspace{1cm}

\caption{Same as Fig.~\ref{ics.s5D} except for the 6s6d $^1$D$_2$  level.
}
\label{ics.s7D}
\end{figure}

Electron impact excitation cross sections for the $5d^2~^1S_0$ level
and $^1D_2$ levels associated with the $6s5d$, $5d^2$,and $6s6d$ major
configurations are given in Figs.~ \ref{ics.s7S}, \ref{ics.s5D},
\ref{ics.s6D}, and \ref{ics.s7D}, respectively.  We did not include
the very approximate $Q$($6s5d~^1D_2$) values of Jensen \etal
\cite{JRT78} in Fig.~\ref{ics.s5D}.  No other results are available
and again, we give our recommended cross sections based on the CCC
calculations in Table~\ref{IV}.

\begin{figure}
\hspace{-2cm}
\epsfbox{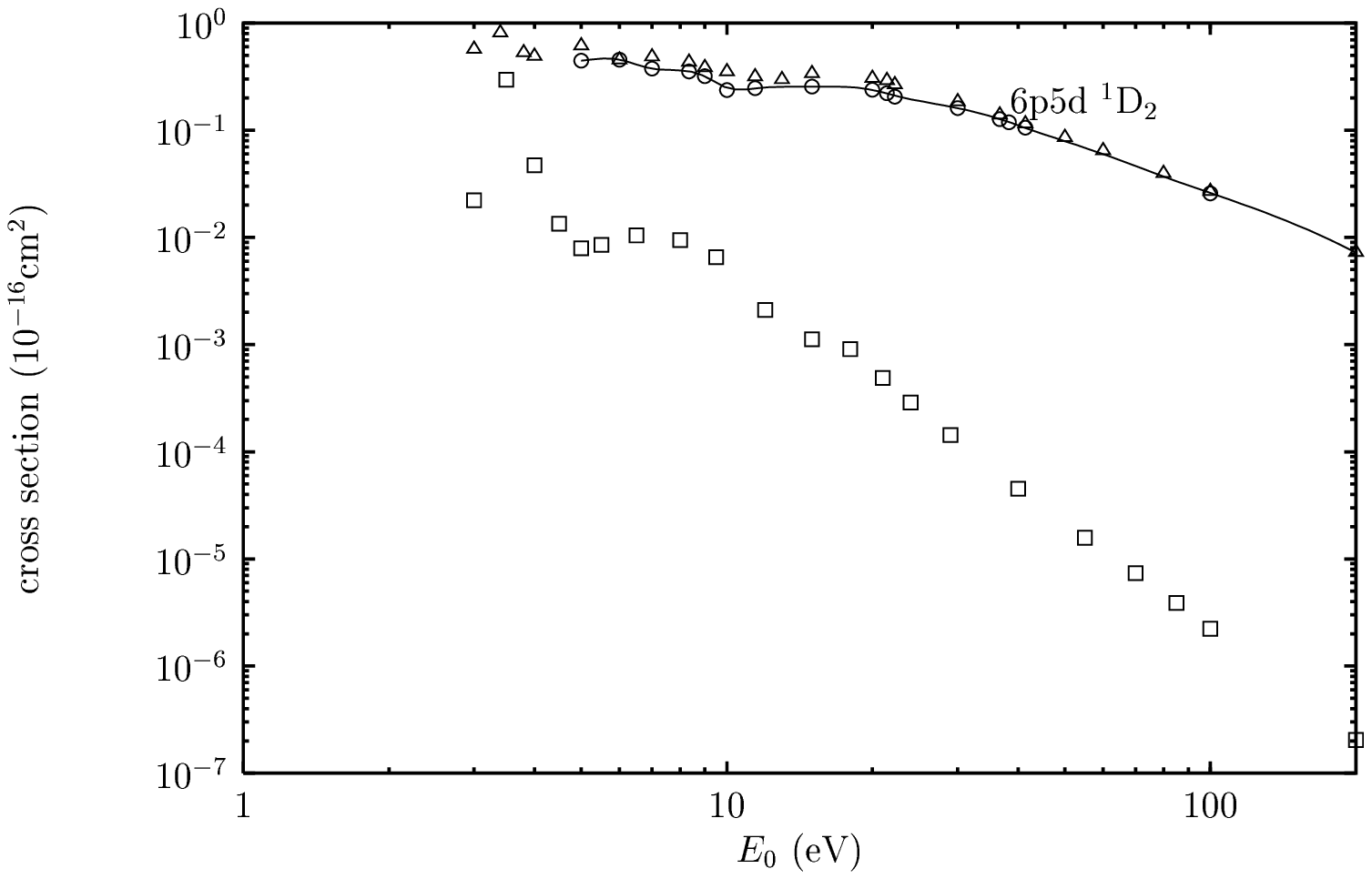}
\vspace{1cm}
\caption{Same as Fig.~\ref{ics.s5D} except for the  6p5d $^1$D$_2$ level.
}
\label{ics.S5D}
\end{figure}

\begin{figure}
\hspace{-2cm}
\epsfbox{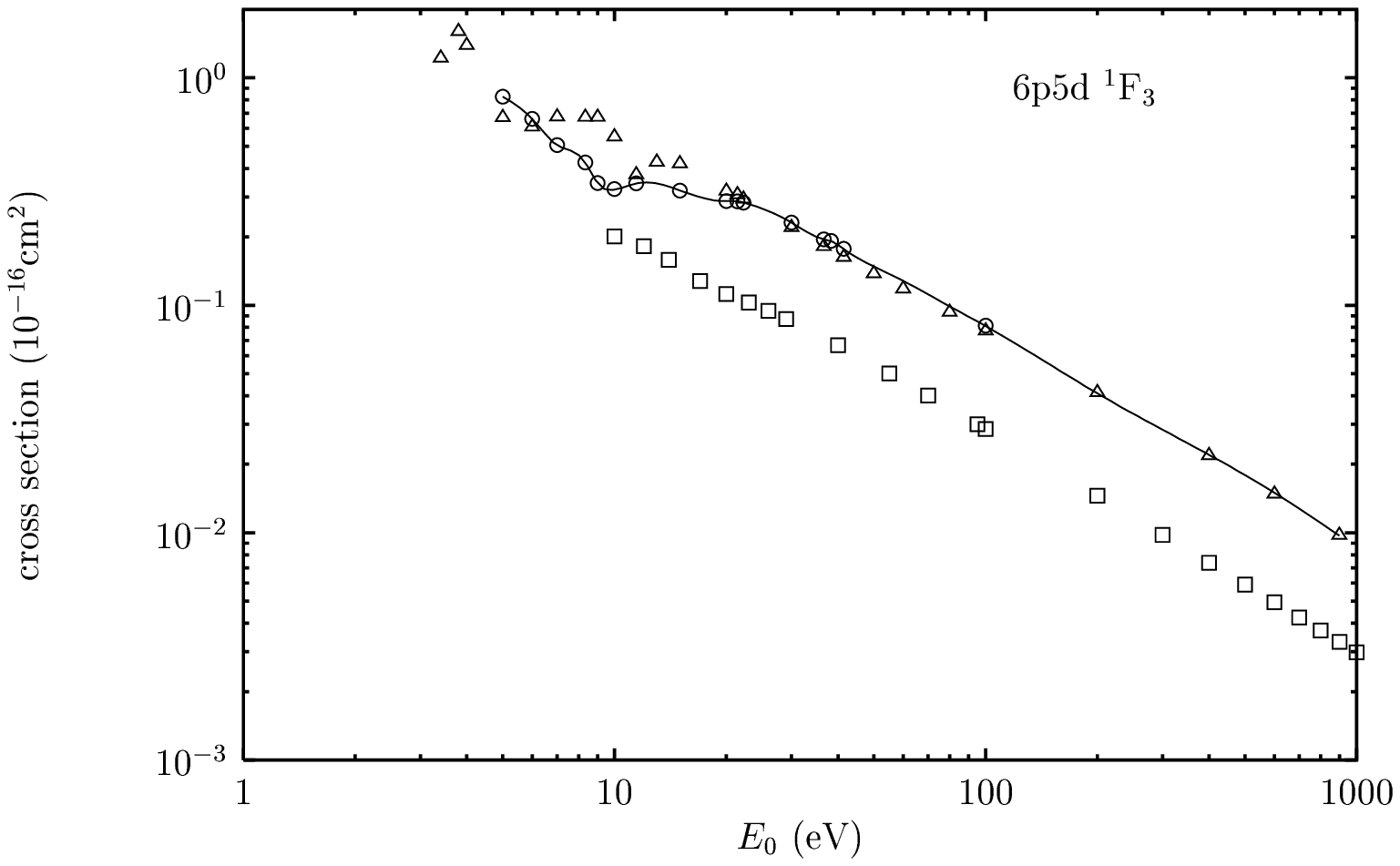}
\vspace{1cm}
\caption{Same as Fig.~\ref{ics.s5D} except for the  6p5d $^1$F$_3$level.
}
\label{ics.s4F}
\end{figure}

\begin{figure}
\hspace{-2cm}
\epsfbox{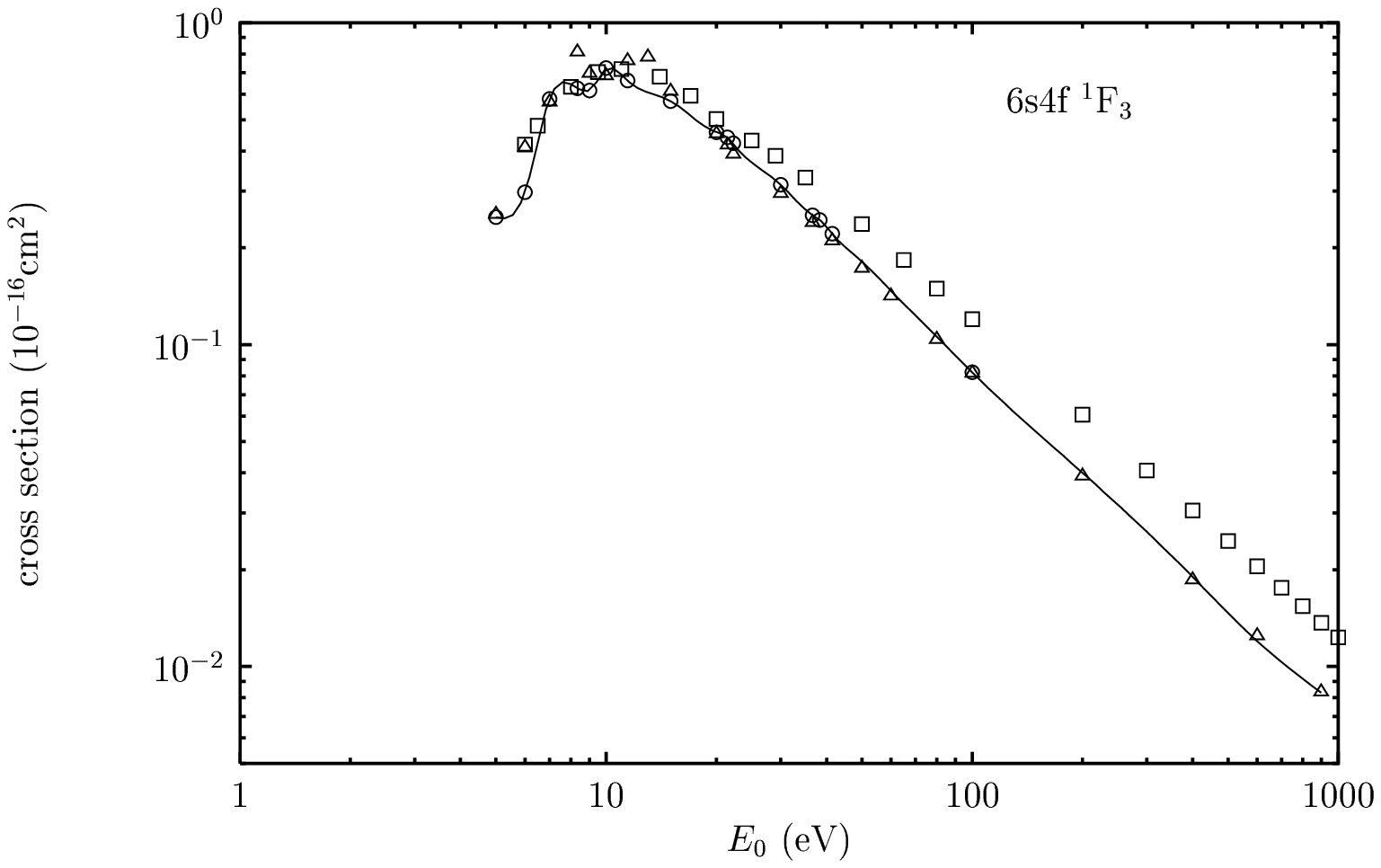}
\vspace{1cm}
\caption{Same as Fig.~\ref{ics.s5D} except for the 6s4f $^1$F$_3$ level.
}
\label{ics.s5F}
\end{figure}

\begin{figure}
\hspace{-2cm}
\epsfbox{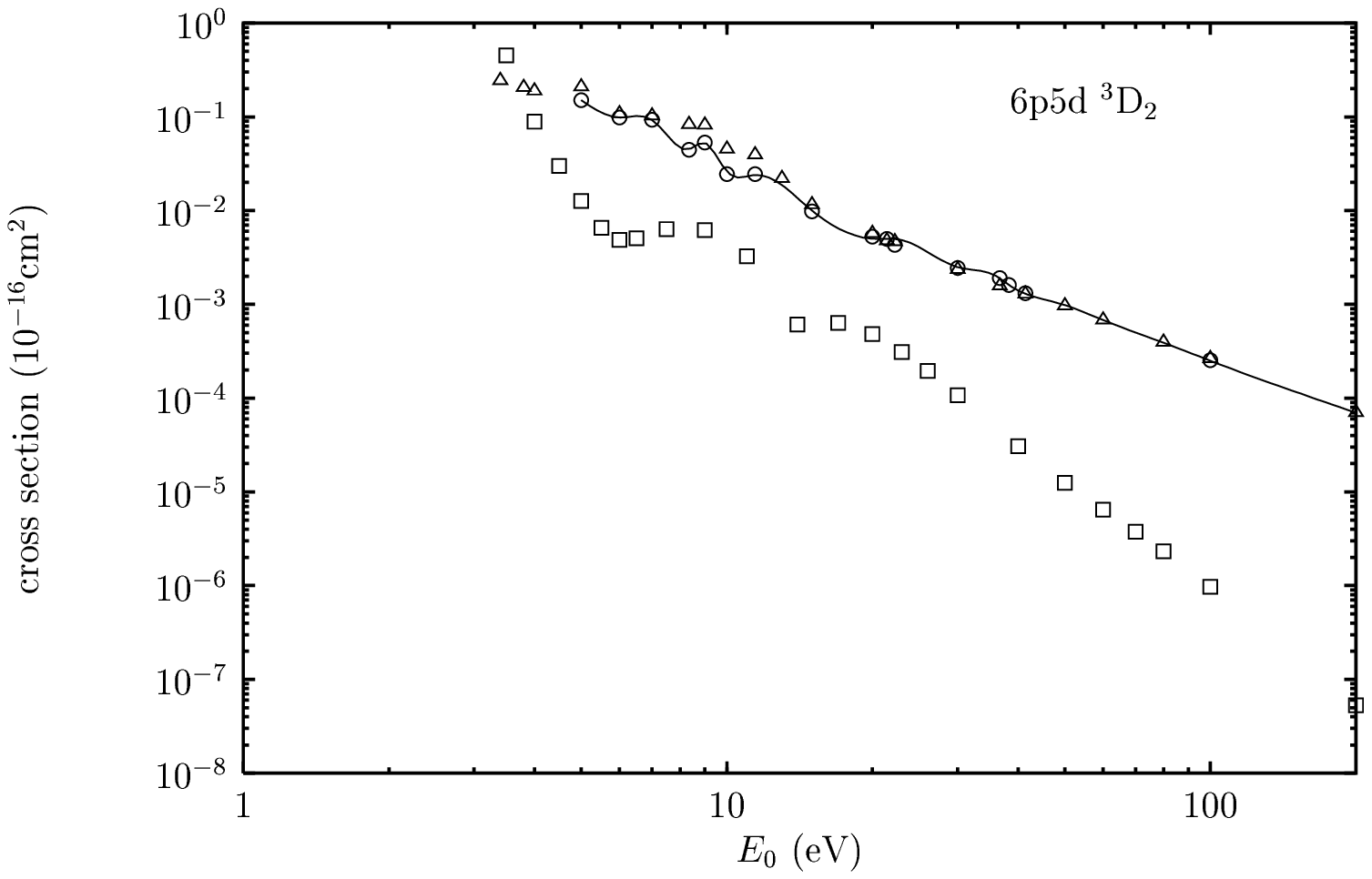}
\vspace{1cm}
\caption{Same as Fig.~\ref{ics.s5D} except for the  6p5d $^3$D$_2$ level.
}
\label{ics.T5D_2}
\end{figure}

Other important excitation channels are associated with the
$6p5d~^1D_2$, $6p5d~^1F_3$, $6s4f~^1F_3$ and $6p5d~^3D_2$ levels.  The
theoretical results for these cross sections are shown in
Figs.~\ref{ics.S5D}, \ref{ics.s4F}, \ref{ics.s5F}, and
\ref{ics.T5D_2}, respectively and the recommended values are listed in
Table~\ref{V}.

\begin{figure}
\hspace{-2cm}
\epsfbox{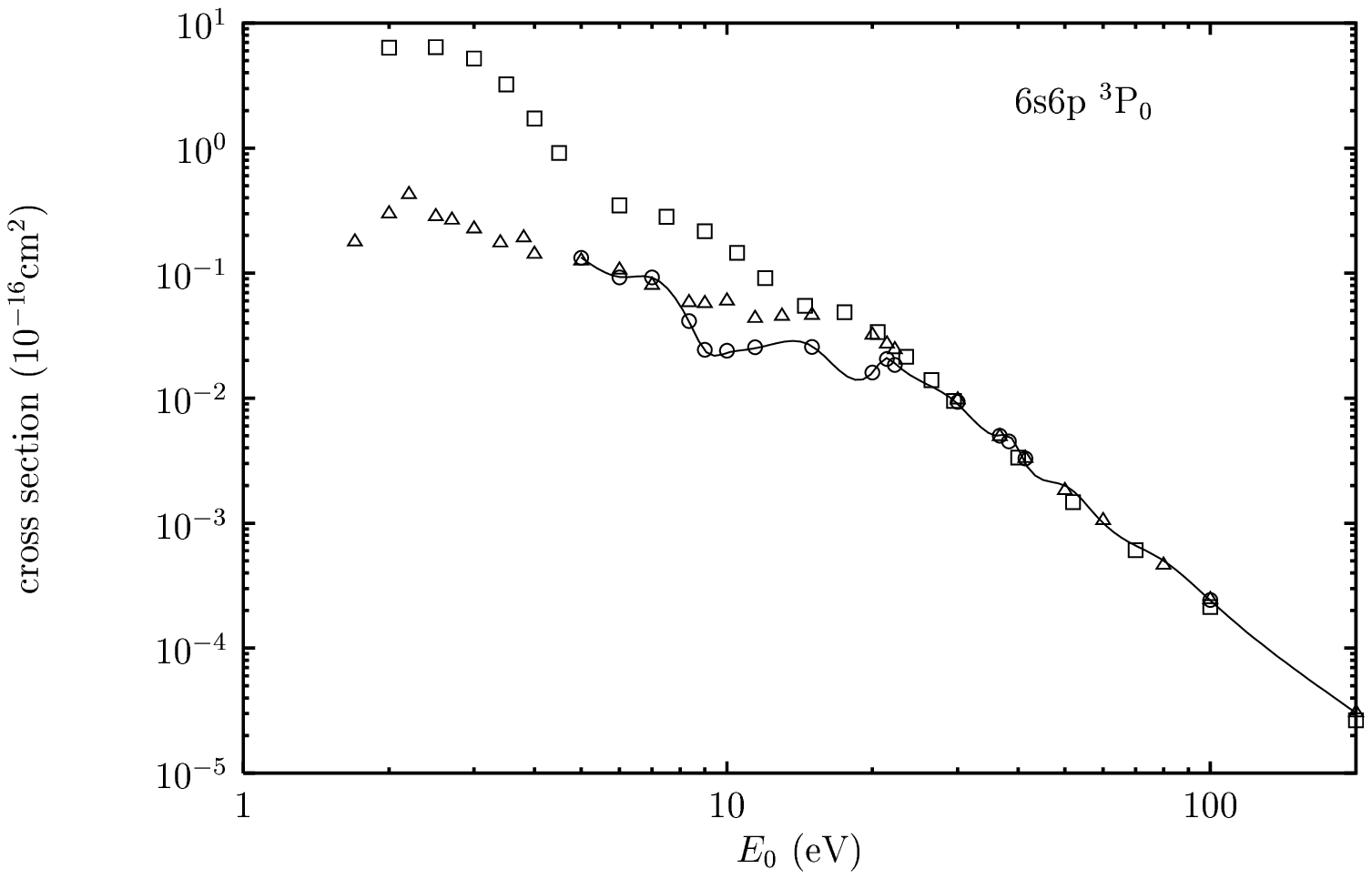}
\vspace{1cm}
\caption{Same as Fig.~\ref{ics.s5D} except for the  6s6p $^3$P$_0$ level.
}
\label{ics.t6P_0}
\end{figure}

\begin{figure}
\hspace{-2cm}
\epsfbox{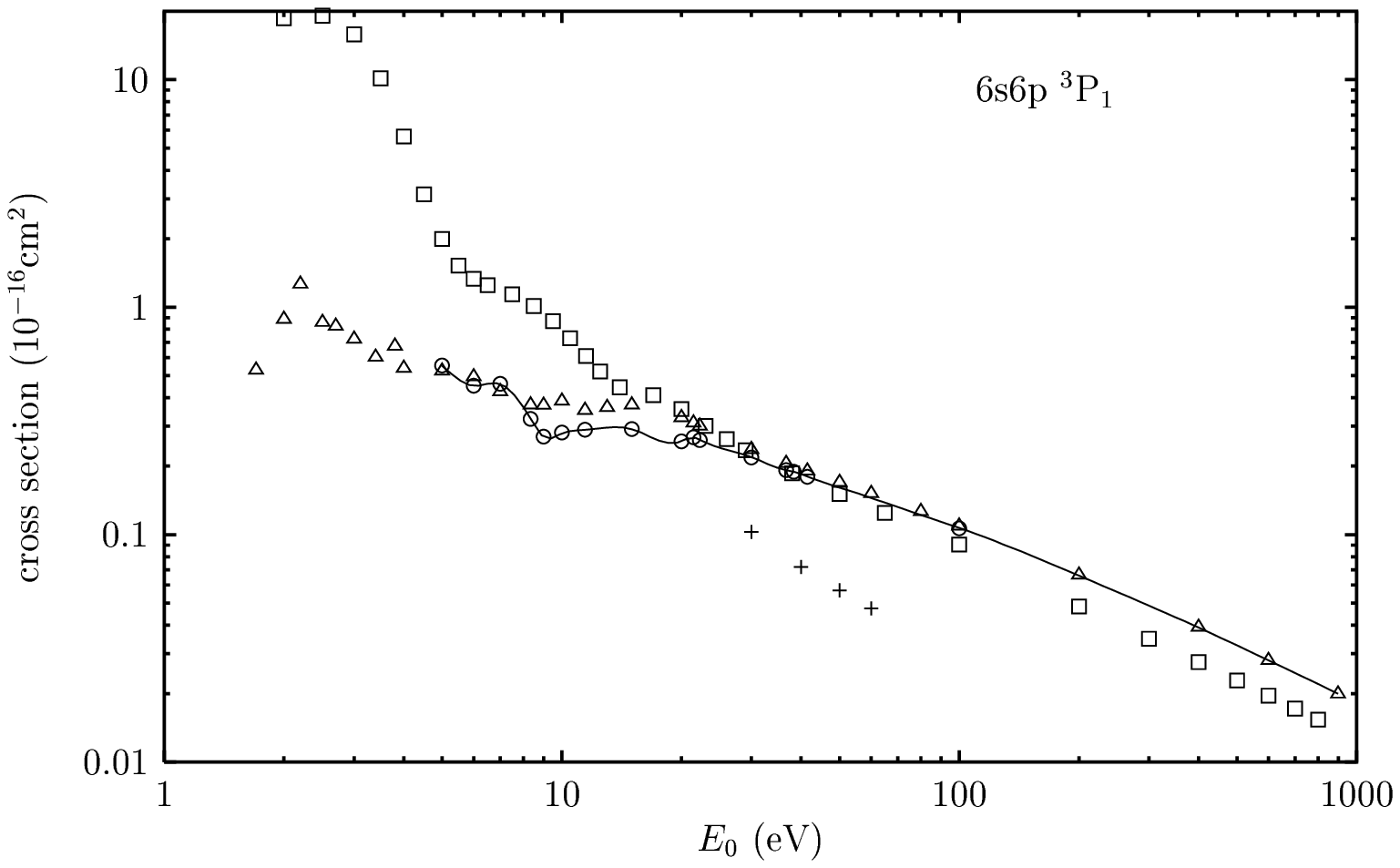}
\vspace{1cm}
\caption{Same as Fig.~\ref{ics.s5D} except for the  6s6p $^3$P$_1$ level.
}
\label{ics.t6P_1}
\end{figure}

\begin{figure}
\hspace{-2cm}
\epsfbox{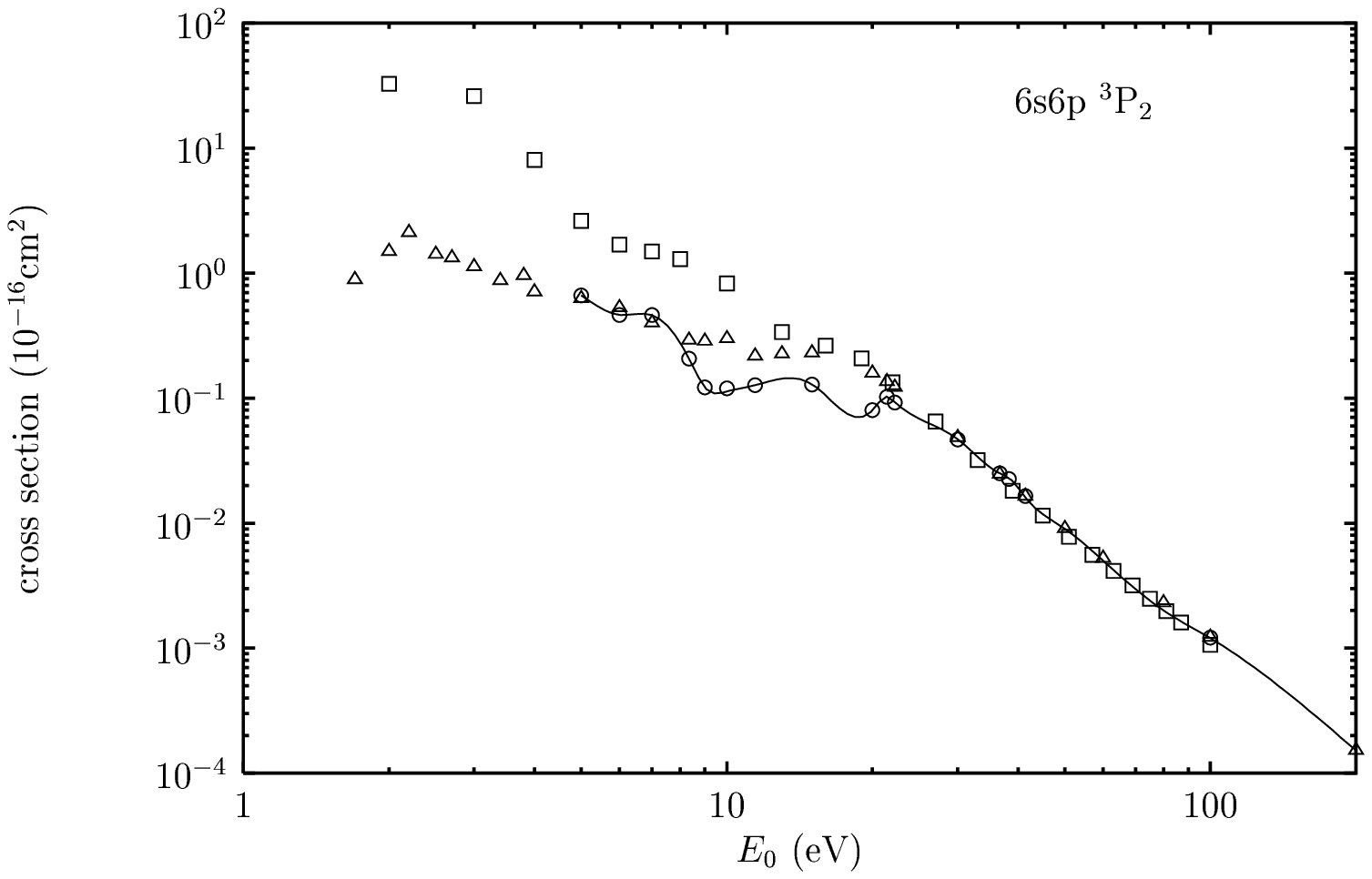}
\vspace{1cm}
\caption{Same as Fig.~\ref{ics.s5D} except for the  6s6p $^3$P$_2$ level.
}
\label{ics.t6P_2}
\end{figure}

\begin{figure}
\hspace{-2cm}
\epsfbox{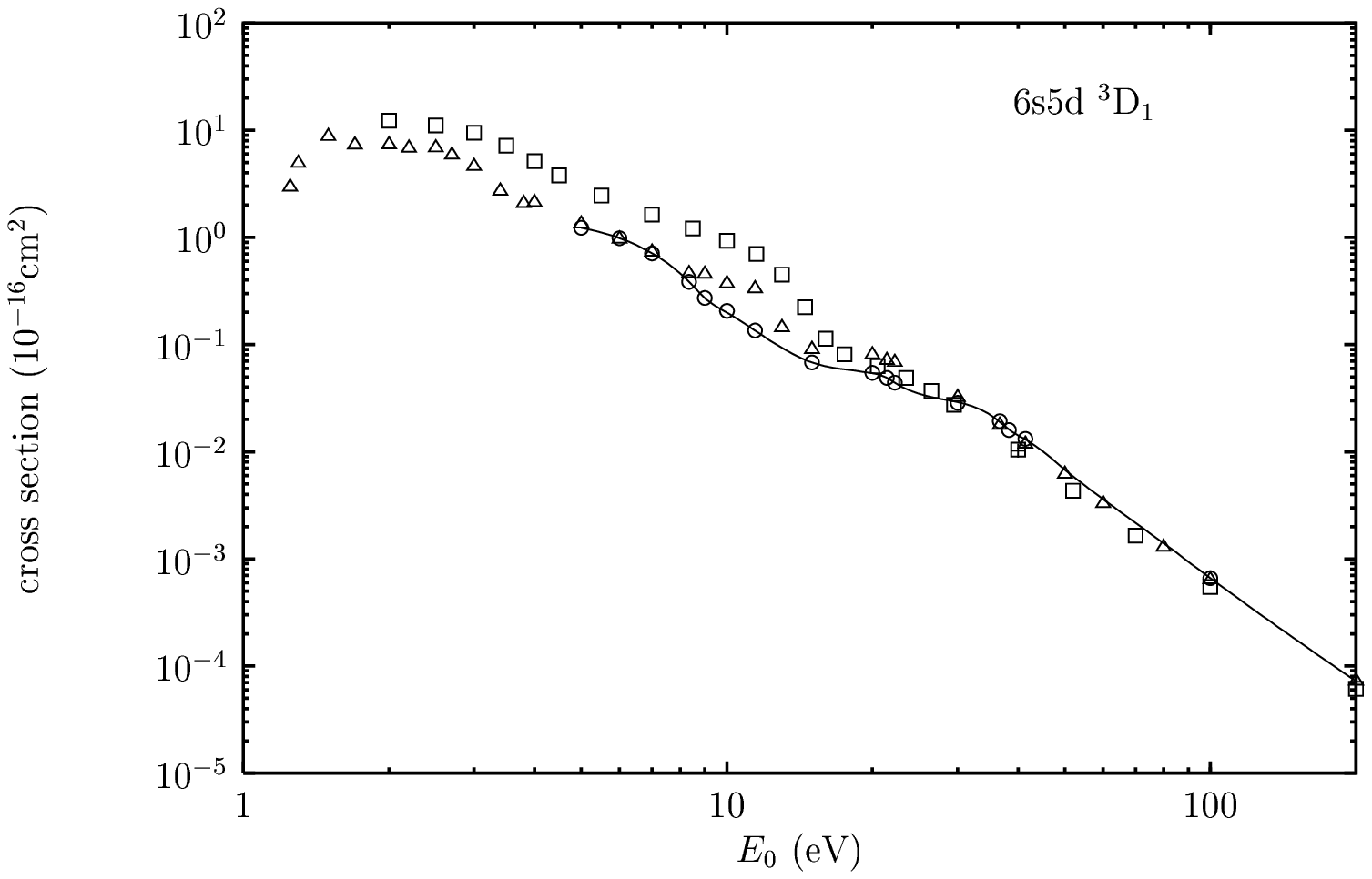}
\vspace{1cm}
\caption{Same as Fig.~\ref{ics.s5D} except for the  6s5d $^3$D$_1$ level.
}
\label{ics.t5D_1}
\end{figure}

\begin{figure}
\hspace{-2cm}
\epsfbox{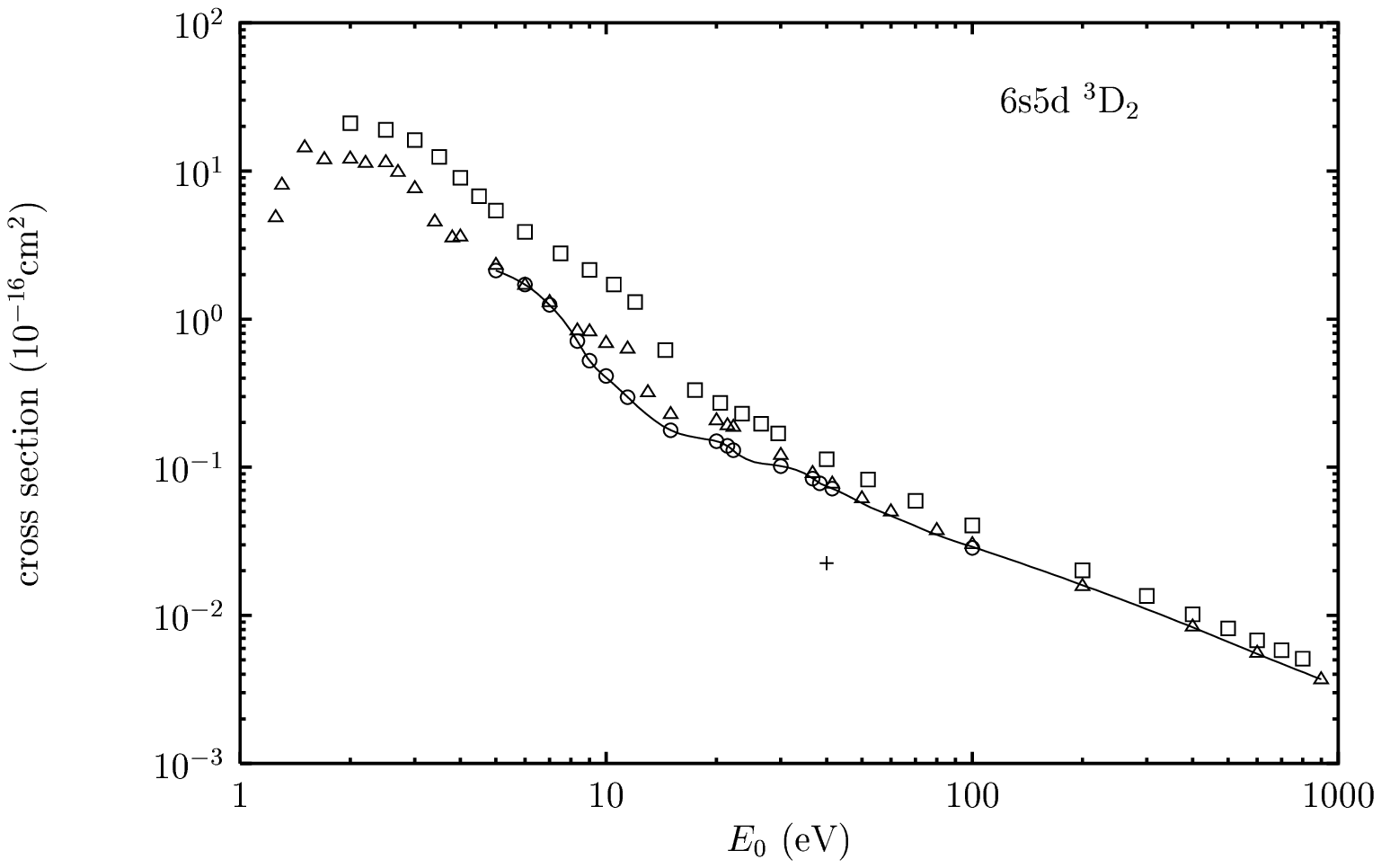}
\vspace{1cm}
\caption{Same as Fig.~\ref{ics.s5D} except for the  6s5d $^3$D$_2$ level.
}
\label{ics.t5D_2}
\end{figure}

\begin{figure}
\hspace{-2cm}
\epsfbox{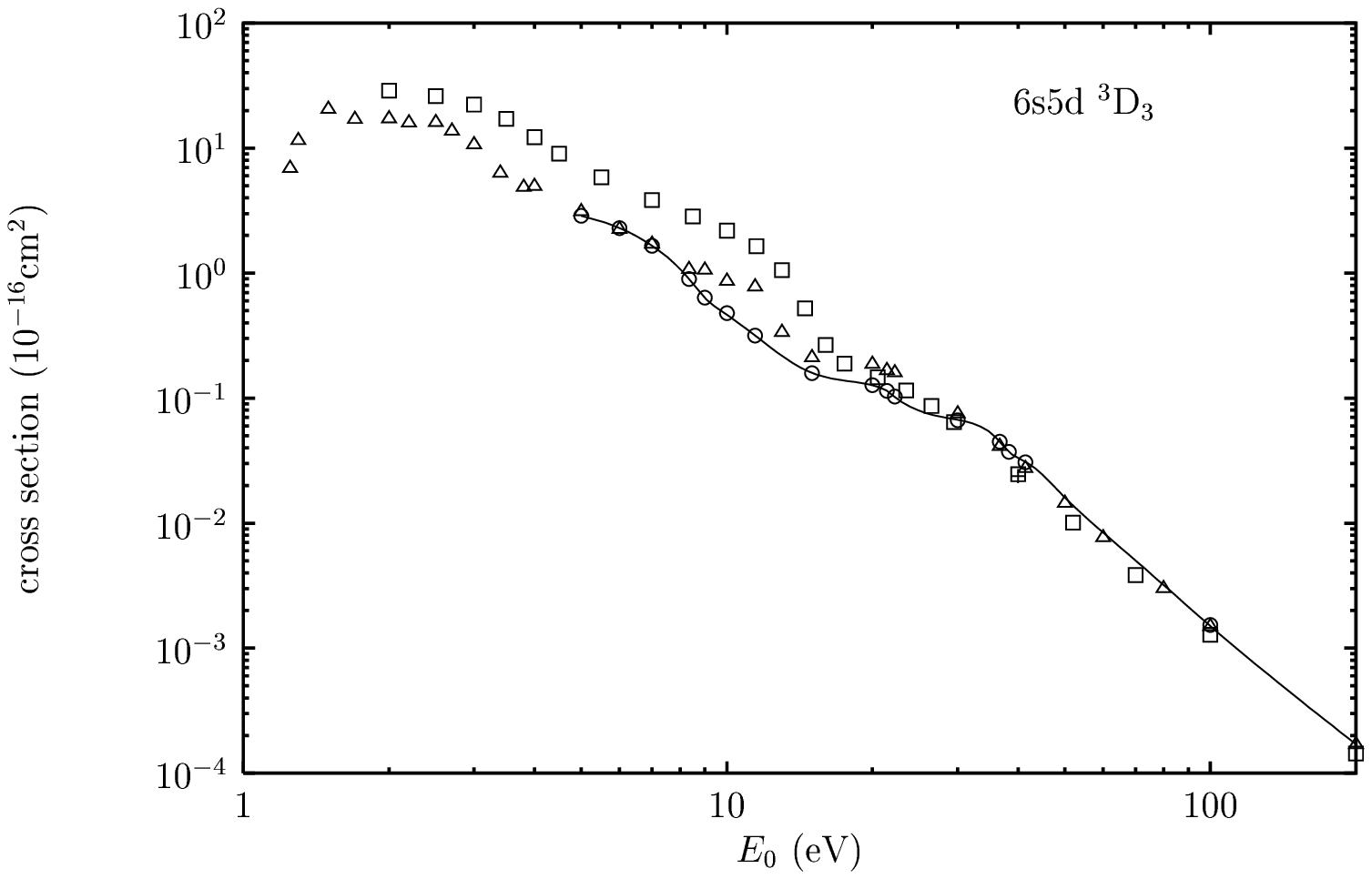}
\vspace{1cm}
\caption{Same as Fig.~\ref{ics.s5D} except for the  6s5d $^3$D$_3$ level.
}
\label{ics.t5D_3}
\end{figure}

Excitation of triplet levels are given for the $6s6p~^3P_J$ ($J=0,1$
and 2), $6s5d~^3D_J$ ($J=1,2$ and 3).  Only theoretical cross sections
are available and they are shown in
Figs.~\ref{ics.t6P_0}-\ref{ics.t6P_2} and
\ref{ics.t5D_1}-\ref{ics.t5D_3}, respectively The recommended values
are summarized in Table~\ref{VI}.

Comparing CCC and UFOMBT results we generally find good agreement at
high incident electron energies. However for a few transitions we
observe substantial discrepancies even at high impact energies.  For
the $6p5d~^1F_3$ state this discrepancy is the result of the small,
but important difference in the CI mixing coefficients for the
$nf6s~^1F_3$ configuration.  We find that the $nf6s~^1F_3$
configuration contributes most to the ICS, specially at high
energies. We gave preference to the CCC results in this case, because
it is likely that the structure calculations performed in the UFOBT
method has not converged for this state. Similarly, for the
$6s5d~^3D_2$ level a small difference in the singlet-triplet mixing
coefficient between $6s5d~^3D_2$ and $6s5d~^1D_2$ configurations leads
to some differences between CCC and UFOMBT calculations at high
energies.

The enormous difference between CCC and UFOMBT results for
$6p5d~^{1}D_2$ and $6p5d~^{3}D_2$ levels has nothing to do with
differences in the structure models but comes from the difference in
the scattering calculations.  In a first order theory, like UFOMBT, in
nonrelativistic approximation the excitation of both $6p5d~^{1}D_2$
and $6p5d~^{3}D_2$ levels from the $6s^2~^1S$ gound state can occur by
exchange scattering only.  As incident electron energy increases, the
exchange scattering decreases which leads to very small values of the
excitation cross section.  Account of relativistic corrections in
UFOMBT does not change this situation because the singlet-triplet
mixing in the ground state is negligible, while the singlet-triplet
mixing for $6p5d~^{1}D_2$ and $6p5d~^{3}D_2$ levels brings
contributions from exchange transitions only.  On other hand, in a
close-coupling theory excitation of $6p5d~^{1}D_2$ level (in
non-relativistic approximation) can occur as a two- (or more) step
processes.  Such processes, for example $6s^2~^1S \rightarrow 6s5d~^{1}D_2
\rightarrow 6p5d~^{1}D_2$, can occur via direct scattering, which
leads to significantly larger cross sections.  The account of
relativistic corrections for the $6p5d~^{3}D_2$ level leads to significant
increase of the cross section due to admixture of the singlet
$6p5d~^{1}D_2$ level, see Eq. (\ref{5h}).

\subsection{Ionization}
\begin{figure}
\hspace{-2cm}
\epsfbox{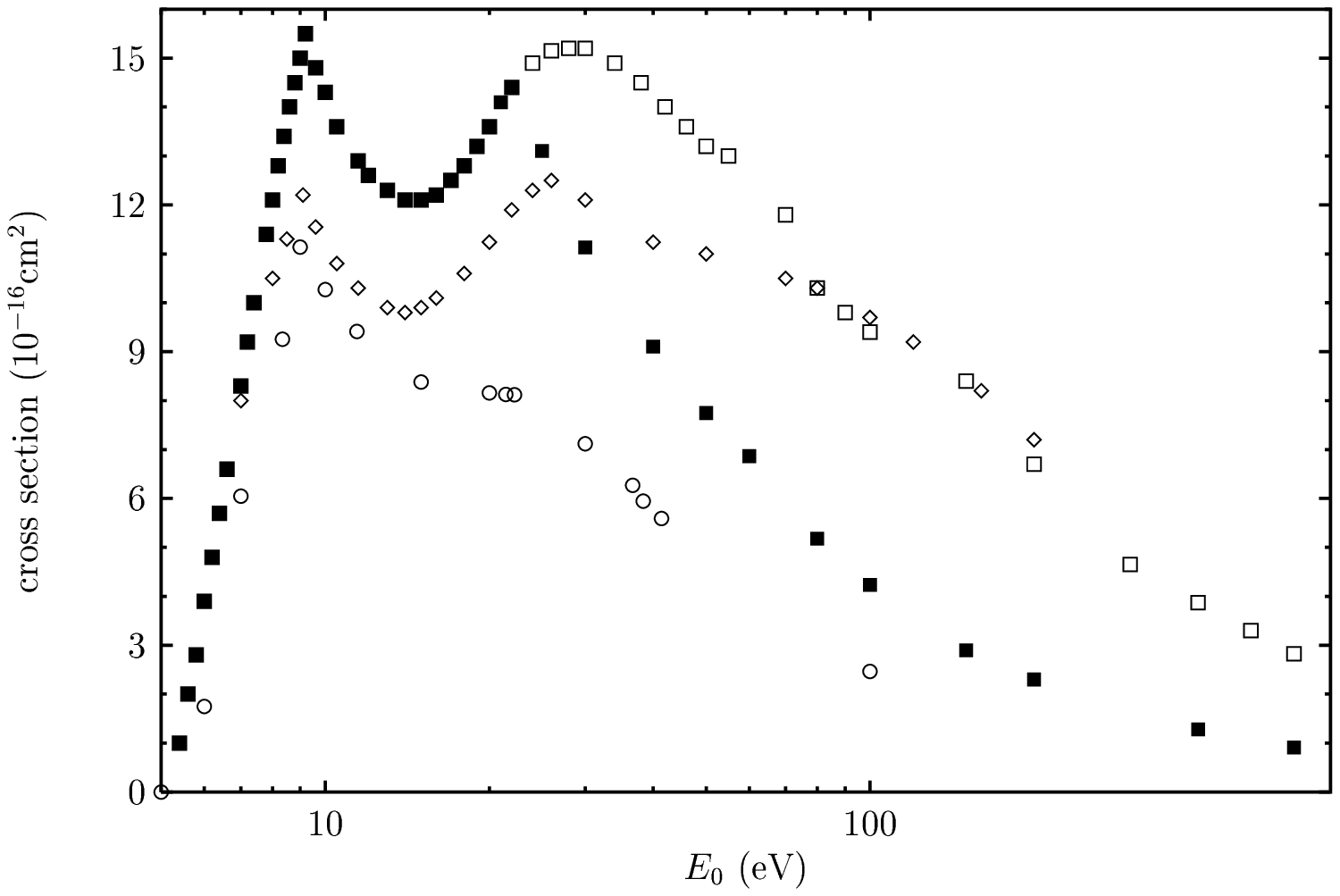}
\vspace{1cm}
\caption{Ionization cross sections: \opencircle, CCC ($Q^+$);
\opensquare, ($Q_i$) and \fullsquare, ($Q^+$)
Dettmann and Karstensen \protect\cite{DK82};
\opendiamond,  ($Q_i$) Vainshtein \etal  \protect\cite{VORS72}.
}
\label{tics}
\end{figure}

Total ionization ($Q^+ + Q^{++} + \dots = Q_i$) and single ionization
($Q^+$) cross sections were measured by Dettmann and Karstensen
\cite{DK82} and total ionization ($Q_i$) by Vainshtein
\etal~\cite{VORS72}.  The CCC results are available only for $Q^+$
(threshold for double ionization is at 15.2 eV). These results are
shown in Fig.~\ref{tics}.  It is clear that the CCC method
substantially underestimates the experimental $Q^+$.  At incident
electron energies above 15 eV this is related to the opening of the
$5p^6$ shell. This process is not accounted for in the CCC model
(which has inert inner shells).  However, below the inner shells
ionization threshold the CCC method should be able to account for all
major ionization channels. Inclusion in the CCC calculations G-states
and other states with larger angular momentum will result in a larger
ionization cross section.  The convergence in the TICS, with
increasing target-space orbital angular momentum, is relatively
fast~\cite{B94l} and we estimate that CCC results should converge to
values 10\%-15\% larger than the present results. This correction of
the CCC results would bring them in a very good agreement with
measurements of TICS by Vainshtein~\etal~\cite{VORS72} in the region
of the first TICS maximum. 
The discrepancy between the experimental
results and between the experimental and the theoretical results in
this energy range makes it impossible to for us to 
present a reliable set of recommended TICS values.
More accurate
theoretical calculations or/and new independent measurements are
required to draw any definite conclusions.  For the time being, we
arbitrarily renormalized the results of Dettmann and Karstensen
\cite{DK82} at the first maximum to the value of 13e-16 cm$^2$.  These
renormalized values are listed in Table~\ref{VII}.

\subsection{Elastic scattering, momentum transfer and total scattering}
\begin{figure}
\hspace{-2cm}
\epsfbox{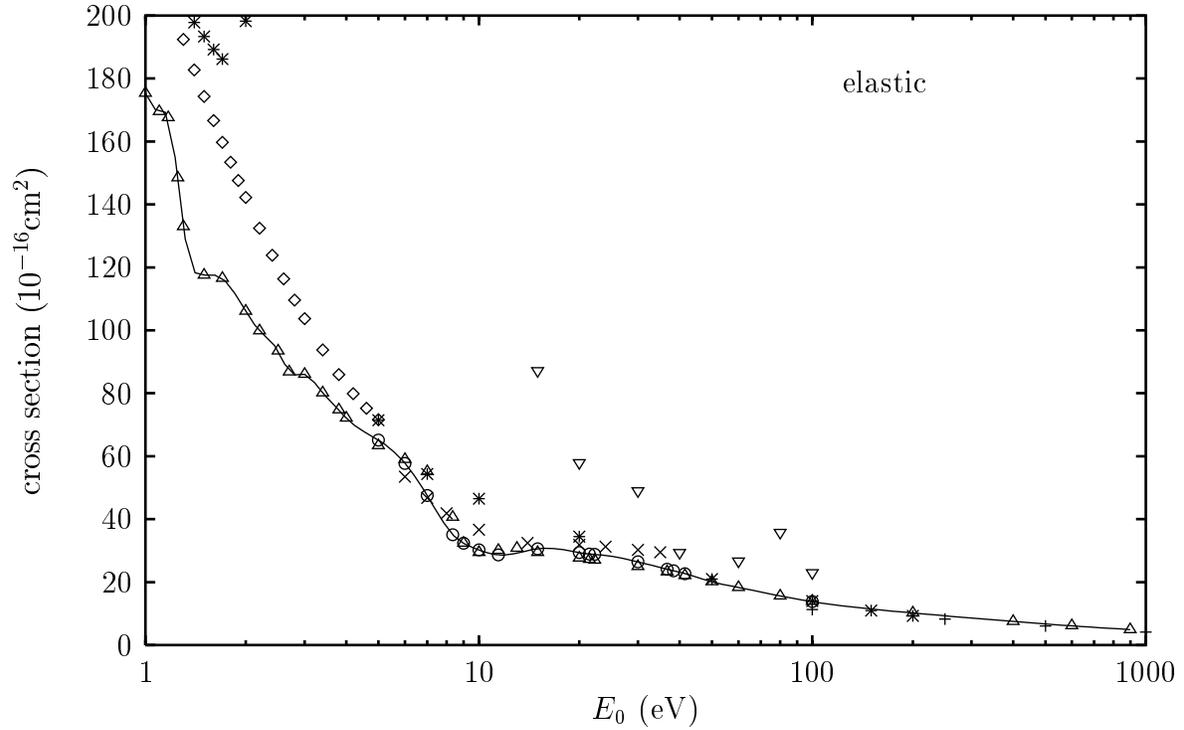}
\vspace{1cm}
\caption{Integral elastic cross sections:
\opencircle,~CCC; \opentriangle,~CC(55); +, Gregory and Fink ~\protect\cite{GF74};
$\times$,~CC(2) Fabrikant \protect\cite{Fabrikant80};
\opentriangledown, Szmytkowski and Sienkiewicz ~\protect\cite{SS94};
\opendiamond, Yuan and Zhang ~\protect\cite{YZ90};
*, Kelemen \etal ~\protect\cite{KRS95}. 
The solid line represents our recommended values.
}
\label{ics.s6S}
\end{figure}

\begin{figure}
\hspace{-2cm}
\epsfbox{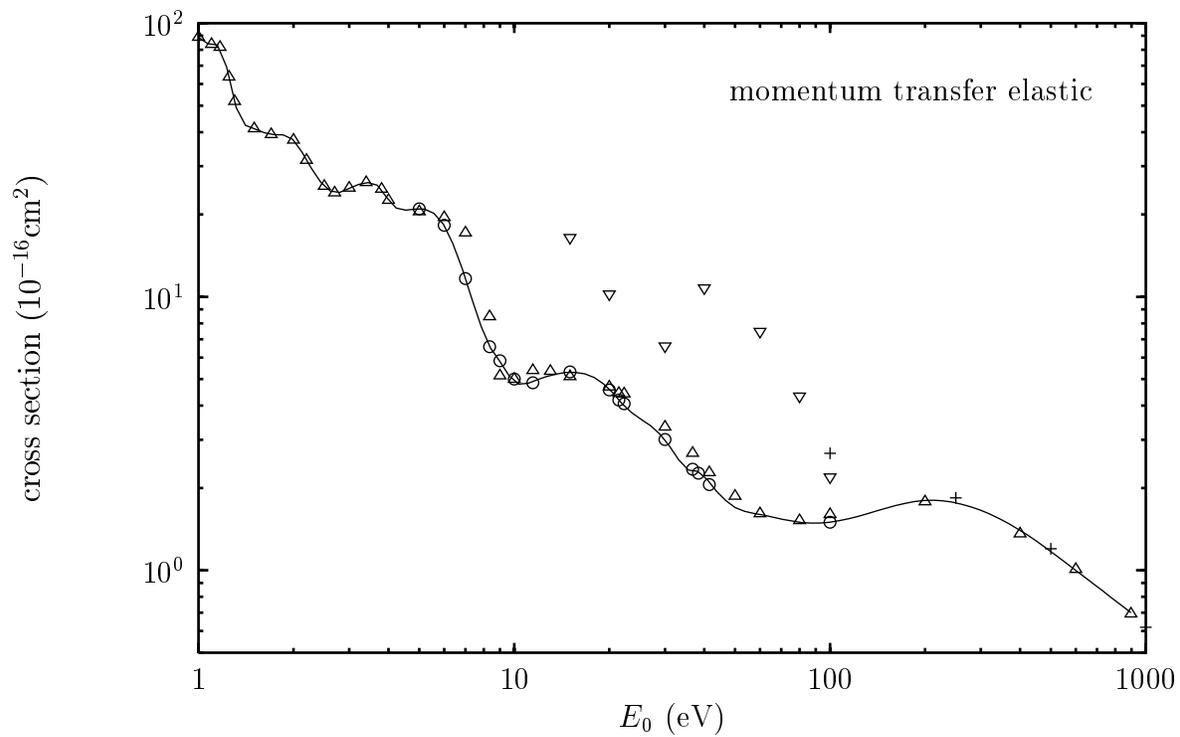}
\vspace{1cm}
\caption{Same as for Fig.~\ref{ics.s6S} but for momentum transfer
cross sections. 
}
\label{ics.s6S.mtr}
\end{figure}

\begin{figure}
\hspace{-2cm}
\epsfbox{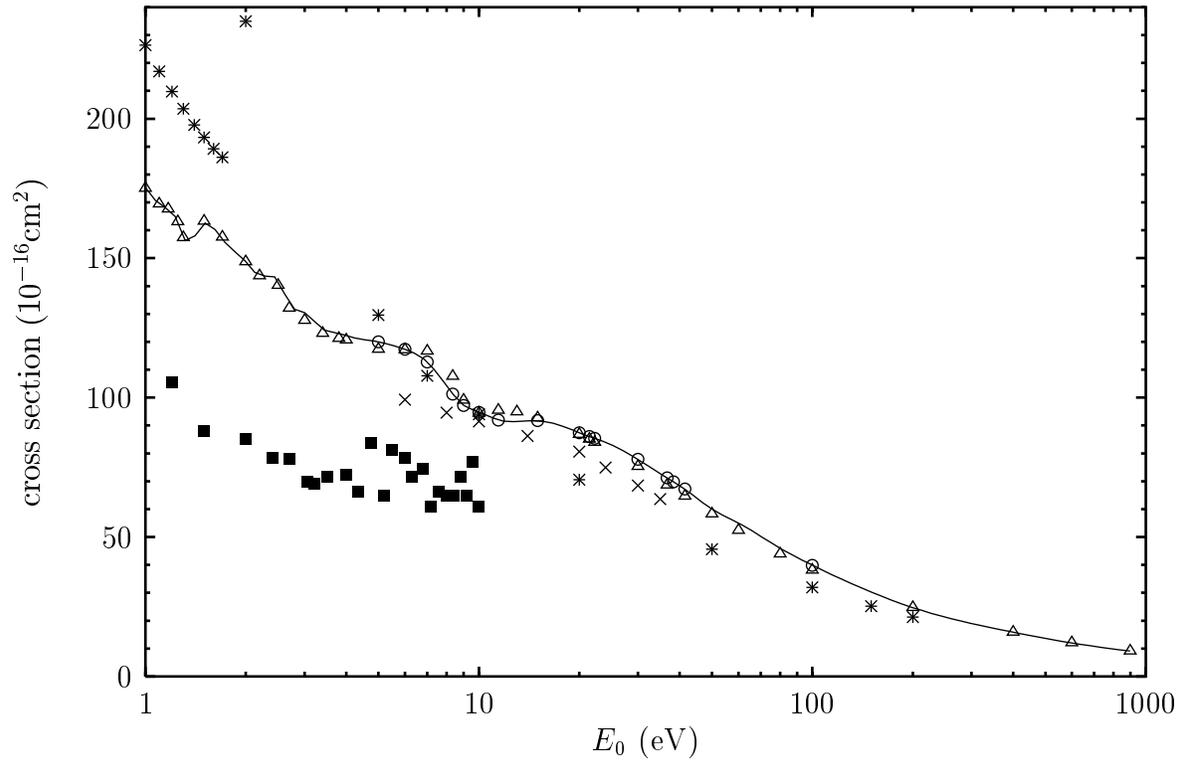}
\vspace{1cm}
\caption{Total electron scattering cross sections:
\opencircle, CCC; \opentriangle, CC(55);
$\times$, CC(2) Fabrikant \protect\cite{Fabrikant80};
*, Kelemen \etal \protect\cite{KRS95};
\fullsquare, Romanyuk \etal \protect\cite{RSZ80}.
The solid line represents our recommended values.}
\label{tcs}
\end{figure}

Elastic scattering and momentum transfer cross sections are available
from a number of calculations. They are shown in Figs.~\ref{ics.s6S}
and \ref{ics.s6S.mtr}, respectively.  Our recommended values are given
in Table~\ref{VIII}, where we have also included the recommeded total
electron scattering cross sections, see Fig.~\ref{tcs}, based mainly
on the CCC results.  At low energies, the experimental results of Romanyuk
\etal \cite{RSZ80} are in poor agreement with our results as well as
with the results of all other calculations. Hence we suppose that the present
theoretical results are more accurate than the experimental ones.

\section{Conclusions}
We have presented a recommended set of integrated cross sections for
electron scattering by the ground state of barium. For most of the
transitions presented here no previous experimental or theoretical data
are available.  We expect our results to be useful in
practical applications and will stimulate further experimental and
theoretical effort to further improve the cross section data set.

\acknowledgments We are grateful to V.Kelemen and A.Stauffer for
communicating their data in electronic form.  Support of the
Australian Research Council and the Flinders University of South
Australia, the National Science Foundation, and the National
Aeronautic and Space Administration is acknowledged.  We are also
indebted to the South Australian Centre for High Performance Computing
and Communications. The work at the Los Alamos National Laboratory has
been performed under the auspices of the US Department of Energy and
has been partially supported by the Electric Power Research Institute.




\newpage

\begin{table}
\caption{Excitation energies and dominant configurations for the barium levels from CCC and CC(55) non-relativistic calculations. 
The experimental data are from Refs.~\protect\cite{M58} and 
\protect\cite{Pal76} (5d$^2$ $^1$S level). 
States are labeled by the major configuration.
}
\begin{tabular}{lc|lcl} 
     \multicolumn{2}{c|}{experiment} & \multicolumn{3}{c}{present}    \\
    label        & E(eV)& label  & E(eV)& Dominant configurations  \\ \hline
    6s$^2$ $^1$S & 0.00 & 6s$^2$ & 0.00 &  0.944(6s$^2$ $^1$S) +  0.228(6p$^2$ $^1$S) - 0.191(7s6s $^1$S)  \\ 
    5d$^2$ $^1$S & 3.32 & 5d$^2$ & 3.34 &  0.591(7s6s $^1$S) - 0.519(5d$^2$ $^1$S) + 0.369(nd5d $^1$S) \\ 
    6s6p $^1$P   & 2.24 & 6s6p   & 2.27 &  0.800(6p6s $^1$P) - 0.504(5d6p $^1$P)-  0.256(7p6s $^1$P) \\ 
    6s7p $^1$P   & 3.54 & 6s7p   & 3.62 &  0.688(7p6s $^1$P) - 0.550(5d6p $^1$P) + 0.331(5d7p $^1$P) \\ 
    6s8p $^1$P   & 4.04 & 6s8p   & 4.14 &  0.788(6snp $^1$P)  + 0.301(5d6p $^1$P) - 0.505(5d7p $^1$P) \\ 
    6s5d $^1$D   & 1.41 & 6s5d   & 1.44 &  0.896(5d6s $^1$D) - 0.226(5d7s $^1$D) - 0.226(5d$^2$ $^1$D)\\ 
    5d$^2$ $^1$D & 2.86 & 5d$^2$ & 3.04 &  0.798(5d$^2$ $^1$D) - 0.442(nd5d $^1$D) + 0.350(6p$^2$ $^1$D) \\ 
    6s6d $^1$D   & 3.75 & 6s6d   & 3.79 &    0.893(nd6s $^1$D) - 0.369(5d7s $^1$D) - 0.162(6p$^2$ $^1$D) \\ 
    5d6p $^1$D   & 2.86 & 5d6p   & 2.87 &  0.946(5d6p $^1$D) - 0.289(5d7p $^1$D)\\ 
    5d6p $^1$F   & 3.32 & 5d6p   & 3.35 &  0.852(5d6p $^1$F) - 0.424(5d7p $^1$F) + 0.280(nf6s  $^1$F )\\ 
    6s4f  $^1$F  & 4.31 & 6s4f   & 4.36 &  0.973(nf6s  $^1$F ) + 0.165(5d7p $^1$F) - 0.141(nd6p $^1$F)  \\ 
    6s6p $^3$P   & 1.62 & 6s6p   & 1.59 &  0.960(6p6s $^3$P) - 0.161(5d6p $^3$P) - 0.116(6p7s $^3$P) \\ 
   5d6p $^3$P    & 3.20 & 5d6p   & 3.30 &  0.873(5d6p $^3$P) - 0.394(5d7p $^3$D) - 0.215(7p6s $^3$P) \\ 
   5d$^2$ $^3$P  & 2.94 & 5d$^2$ & 3.11 &  0.799(5d$^2$ $^3$P) + 0.458(nd5d $^3$D) + 0.389(6p$^2$ $^3$P) \\ 
     6s5d $^3$D  & 1.16 & 6s5d   & 1.21 &  0.955(5d6s $^3$D) - 0.201(5d7s $^3$D)- 0.112(nf6p $^3$D) \\ 
     6s6d $^3$D  & 3.85 & 6s6d   & 3.82 &  0.961(nd6s $^3$D) - 0.208(5d7s $^3$D)\\ 
    5d6p $^3$D   & 3.06 & 5d6p   & 3.12 &  0.924(5d6p $^3$D) - 0.361(5d7p $^3$D) \\ 
    5d6p $^3$F   & 2.86 & 5d6p   & 2.88 &  0.934(5d6p $^3$F) - 0.289(5d7p $^3$F) + 0.129(nf6s  $^3$F ) 
\end{tabular}
\label{energy.levels}
\end{table}

\clearpage

\begin{table}
\caption{Recommended values for $Q_{app}$(6s6p$^1$P$_1$),
$Q_{\rm cascade}$(6s6p$^1$P$_1$), and $Q$(6s6p$^1$P$_1$) in units of $10^{-16}$
cm$^2$.
}
\begin{tabular}{|d|d|d|d|d|} 
 $E_0 \;$ (eV) & $Q_{\rm App}$(6s6p$^1$P$_1$) & $Q_{\rm cascade} / Q_{\rm App} \; (\%) $ & $Q_{\rm cascade}$(6s6p$^1$P$_1$) & $Q$(6s6p$^1$P$_1$)  \\  \hline
2.50 & 4.56 & 0.00 & 0.00 & 4.56 \\
3.00 & 12.00 & 0.00 & 0.00 & 12.00 \\
4.00 & 25.84 & 15.54 & 4.02 & 21.83 \\
5.00 & 33.34 & 17.71 & 5.90 & 27.43 \\
6.00 & 37.26 & 19.52 & 7.27 & 29.98 \\
7.00 & 39.89 & 20.61 & 8.22 & 31.67 \\
8.35 & 39.00 & 19.80 & 7.72 & 31.28 \\
9.00 & 40.44 & 17.53 & 7.09 & 33.35 \\
10.00 & 41.24 & 15.63 & 6.45 & 34.79 \\
11.44 & 42.56 & 13.48 & 5.74 & 36.82 \\
15.00 & 42.47 & 15.23 & 6.47 & 36.00 \\
20.00 & 39.78 & 13.14 & 5.23 & 34.55 \\
30.00 & 35.01 & 11.19 & 3.92 & 31.09 \\
36.67 & 32.39 & 10.38 & 3.36 & 29.03 \\
41.44 & 30.78 & 9.94 & 3.06 & 27.72 \\
50.00 & 28.11 & 9.57 & 2.69 & 25.42 \\
60.00 & 25.49 & 9.15 & 2.33 & 23.16 \\
80.00 & 21.55 & 8.30 & 1.79 & 19.76 \\
100.00 & 18.75 & 7.44 & 1.40 & 17.35 \\
200.00 & 11.65 & 6.56 & 0.76 & 10.88 \\
400.00 & 6.81 & 5.46 & 0.37 & 6.44 \\
600.00 & 4.92 & 5.17 & 0.25 & 4.66 \\
897.60 & 3.52 & 4.91 & 0.17 & 3.35 \\
\end{tabular}
\label{II}
\end{table}

\clearpage

\begin{table}
\caption{Recommended $Q$(6s7p$^1$P$_1$), $Q_{app}$(6s7p$^1$P$_1$), $Q$(6s8p$^1$P$_1$), 
and $Q_{app}$(6s8p$^1$P$_1$) values in  units of $10^{-16}$ cm$^2$.
}
\begin{tabular}{|d|d|d|d|d|} 
 $E_0 \;$ (eV) & $Q$(6s7p$^1$P$_1$) &  $Q_{\rm App}$(6s7p$^1$P$_1$)  & $Q$(6s8p$^1$P$_1$) & $Q_{\rm App}$(6s8p$^1$P$_1$)  \\  \hline
5.00 &0.48 &0.70 &0.22 &0.27 \\
6.00 &0.76 &1.04 &0.45 &0.53 \\
7.00 &0.73 &1.06 &0.71 &0.81 \\
8.35 &0.81 &1.20 &0.78 &0.90 \\
9.00 &0.49 &0.86 &0.64 &0.76 \\
10.00 &0.35 &0.70 &0.69 &0.80 \\
11.44 &0.33 &0.60 &1.00 &1.08 \\
15.00 &0.32 &0.65 &1.18 &1.30 \\
20.00 &0.39 &0.67 &1.30 &1.40 \\
30.00 &0.47 &0.71 &1.45 &1.54 \\
36.67 &0.50 &0.72 &1.46 &1.54 \\
50.00 &0.50 &0.70 &1.49 &1.55 \\
60.00 &0.49 &0.65 &1.45 &1.50 \\
80.00 &0.46 &0.58 &1.25 &1.35 \\
100.00 &0.42 &0.54 &1.10 &1.14 \\
200.00 &0.30 &0.37 &0.74 &0.77 \\
400.00 &0.19 &0.23 &0.44 &0.45 \\
600.00 &0.14 &0.17 &0.31 &0.32 \\
897.60 &0.10 &0.12 &0.23 &0.23 \\
\end{tabular}
\label{III}
\end{table}

\clearpage

\begin{table}
\caption{Recommended $Q$(5d$^2$~$^1$S$_0$), $Q$(6s5d$^1$D$_2$), $Q$(5d$^2$~$^1$D$_2$), 
and $Q$(6s6d$^1$D$_2$) values in units of $10^{-16}$ cm$^2$.
}
\begin{tabular}{|d|d|d|d|d|} 
 $E_0 \;$ (eV) & $Q$(5d$^2$~$^1$S$_0$) &  $Q$(6s5d$^1$D$_2$) &   $Q$(5d$^2$~$^1$D$_2$) & $Q$(6s6d$^1$D$_2$)  \\  \hline
5.00 &0.81 &5.45 &2.74 &1.21 \\
6.00 &0.89 &4.95 &2.54 &1.79 \\
7.00 &1.23 &4.07 &2.41 &2.25 \\
8.35 &1.79 &3.69 &1.87 &2.21 \\
9.00 &1.17 &3.59 &1.57 &1.96 \\
10.00 &0.70 &3.53 &1.51 &2.01 \\
11.44 &0.55 &3.39 &1.44 &2.28 \\
15.00 &0.48 &2.99 &1.28 &2.13 \\
20.00 &0.36 &2.74 &0.90 &1.89 \\
30.00 &0.38 &2.46 &0.51 &1.50 \\
36.67 &0.38 &2.33 &0.37 &1.29 \\
41.44 &0.38 &2.24 &0.31 &1.15 \\
50.00 &0.35 &2.00 &0.24 &0.97 \\
60.00 &0.31 &1.78 &0.19 &0.81 \\
80.00 &0.27 &1.44 &0.13 &0.62 \\
100.00 &0.23 &1.22 &0.10 &0.50 \\
200.00 &0.14 &0.69 &0.043 &0.25 \\
400.00 &0.08 &0.37 &0.018 &0.125 \\
600.00 &0.05 &0.25 &0.012&0.083 \\
897.60 &0.03 &0.16 &0.008 &0.056 \\
\end{tabular}
\label{IV}
\end{table}

\clearpage

\begin{table}
\caption{Recommended $Q$(6p5d$^1$D$_2$), $Q$(6p5d$^1$F$_3$), $Q$(6s4f$^1$F$_3$), 
and $Q$(6p5d$^3$D$_2$) values in units of $10^{-16}$ cm$^2$.
}
\begin{tabular}{|d|d|d|d|d|} 
 $E_0 \;$ (eV) & $Q$(6p5d$^1$D$_2$) & $Q$(6p5d$^1$F$_3$)  &  $Q$(6s4f$^1$F$_3$) &  $Q$(6p5d$^3$D$_2$) \\  \hline
5.00 &0.446 &0.826 &0.249 &0.150 \\
6.00 &0.456 &0.661 &0.297 &0.098 \\
7.00 &0.376 &0.506 &0.580 &0.093 \\
8.35 &0.355 &0.424 &0.626 &0.045 \\
9.00 &0.319 &0.345 &0.617 &0.053 \\
10.00 &0.249 &0.322 &0.718 &0.026 \\
11.44 &0.246 &0.344 &0.661 &0.024 \\
15.00 &0.256 &0.320 &0.571 &0.010 \\
20.00 &0.238 &0.287 &0.457 &0.005 \\
30.00 &0.161 &0.231 &0.314 &0.0025 \\
36.67 &0.127 &0.195 &0.252 &0.0019 \\
41.44 &0.105 &0.177 &0.221 &0.0013 \\
50.00 &0.079 &0.148 &0.181 &0.00098 \\
60.00 &0.060 &0.128 &0.147 &0.00068 \\
80.00 &0.037 &0.099 &0.106 &0.00039 \\
100.00 &0.026 &0.081 &0.082 &0.00025 \\
200.00 &0.0072 &0.041 &0.04 & - \\
400.00 &0.0009 &0.022 &0.019 & - \\
600.00 &0.00025 &0.015 &0.012 & - \\
897.60 &0.00017 &0.0097 &0.0083 & - \\
\end{tabular}
\label{V}
\end{table}

\clearpage

\begin{table}
\caption{Recommended $Q$(6s6p$^3$P$_J$) and  $Q$(6s5d$^3$D$_J$) 
values in units of $10^{-16}$ cm$^2$.
}
\begin{tabular}{|d|d|d|d|d|d|d|} 
 $E_0 \;$ (eV) & \multicolumn{3}{c|}{ $Q$(6s6p$^3$P$_J$) } &  \multicolumn{3}{c}{ $Q$(6s5d$^3$D$_J$) } \\  \hline
               & J=0  & J=1  & J=2 &  J=1  & J=2  & J=3  \\ \hline
5.00 &0.133 &0.553 &0.664 &1.232 &2.130 &2.875 \\
6.00 &0.093 &0.451 &0.463 &0.983 &1.712 &2.293 \\
7.00 &0.092 &0.460 &0.461 &0.710 &1.247 &1.656 \\
8.35 &0.041 &0.323 &0.207 &0.385 &0.710 &0.899 \\
9.00 &0.024 &0.269 &0.122 &0.272 &0.524 &0.635 \\
10.00 &0.023 &0.278 &0.113 &0.199 &0.404 &0.464 \\
11.44 &0.025 &0.289 &0.127 &0.135 &0.297 &0.316 \\
15.00 &0.026 &0.291 &0.129 &0.068 &0.178 &0.159 \\
20.00 &0.016 &0.257 &0.080 &0.054 &0.150 &0.127 \\
30.00 &0.009 &0.219 &0.043 &0.029 &0.102 &0.067 \\
36.67 &0.005 &0.192 &0.025 &0.019 &0.084 &0.045 \\
41.44 &0.003 &0.180 &0.016 &0.013 &0.072 &0.031 \\
50.00 &0.002 &0.161 &0.009 &0.0068 &0.057 &0.016 \\
60.00 &0.001 &0.145 &0.005 &0.0036 &0.047 &0.0084 \\
80.00 &0.0005 &0.122 &0.002 &0.0014 &0.035 &0.0032 \\
100.00 &0.00024 &0.107 &0.0012 &0.0007 &0.029 &0.0015   \\
200.00 &-    &0.066  &-    &-    &0.016    &-  \\
400.00 &-    &0.039  &-    &-    &0.0083   &-  \\
600.00 &-    &0.028  &-    &-    &0.0055    & -  \\
897.60 &-    &0.020  &-    &-    &0.0037   &-  \\
\end{tabular}
\label{VI}
\end{table}

\clearpage

\begin{table}
\caption{Estimate of ionization cross section
$Q_{\rm ion}$ and  $Q^+$ values in units of $10^{-16}$ cm$^2$. 
}
\begin{tabular}{|d|d|d|} 
 $E_0 \;$ (eV) &  $Q_{\rm ion}$ &  $Q^+$  \\  \hline
5.40 &0.8  &0.8  \\
6.00 &3.3  &3.3 \\
7.00 &7.0  &7.00 \\
8.00 &10.1 &10.1 \\
9.00 &12.6 &12.6 \\
10.00 &12.0 &12.0 \\
12.00 &10.6 &10.6 \\
15.00 &10.2 &10.2 \\
20.00 &11.4 &11.4 \\
30.00 &12.8 &9.3 \\
40.00 &12.0 &7.6 \\
50.00 &11.1 &6.5 \\
80.00 &8.6  &4.3 \\
100.00 &7.9 &3.6 \\
150.00 &7.1  &2.4 \\
200.00 &5.6  &1.9 \\
400.00 &3.3 &1.1 \\
600.00 &2.4 &0.8
\end{tabular}
\label{VII}
\end{table}

\clearpage

\begin{table}
\caption{Recommended $Q_{\rm elas}$, $Q_M$, and  $Q_{\rm Tot}$
values in units of $10^{-16}$ cm$^2$.
}
\begin{tabular}{|d|d|d|d|} 
 $E_0 \;$ (eV) &  $Q_{\rm elas}$ &  $Q_M$  &  $Q_{\rm Tot}$  \\  \hline
1.00 &175.3 &88.8 &175.3 \\
1.50 &117.5 &41.1 &162.4 \\
2.00 &106.1 &37.4 &148.7 \\
2.50 &93.4 &25.4 &142.0 \\
3.00 &86.0 &24.9 &130.5 \\
4.00 &72.1 &22.5 &122.2 \\
5.00 &65.1 &21.0 &120.0 \\
6.00 &57.8 &18.2 &117.3 \\
7.00 &47.5 &11.7 &112.8 \\
8.35 &35.0 &6.6 &101.3 \\
9.00 &32.3 &5.8 &97.2 \\
10.00 &30.2 &4.9 &94.8 \\
11.44 &28.6 &4.9 &92.0 \\
15.00 &30.6 &5.3 &91.7 \\
20.00 &29.4 &4.6 &87.4 \\
30.00 &26.4 &3.0 &77.8 \\
41.44 &22.7 &2.1 &67.2 \\
50.00 &20.1 &1.7 &60.0 \\
60.00 &18.3 &1.6 &55.0 \\
80.00 &15.6 &1.5 &46.0 \\
100.00 &13.8 &1.5 &39.9 \\
200.00 &10.2 &1.8 &24.7 \\
400.00 &7.5 &1.4 &15.9 \\
600.00 &6.1 &1.0 &12.0 \\
897.60 &4.9 &0.7 &9.1 \\
\end{tabular}
\label{VIII}
\end{table}

\clearpage

\clearpage

\clearpage

\clearpage

\clearpage

\clearpage

\clearpage

\clearpage

\clearpage

\clearpage

\clearpage

\clearpage

\clearpage

\clearpage

\clearpage

\clearpage

\clearpage

\clearpage

\clearpage

\clearpage

\clearpage

\clearpage

\clearpage

\clearpage

\clearpage

\end{document}